\documentclass{article}

\usepackage{amssymb,amsfonts,amsmath,stmaryrd,longtable}
\usepackage{cite,enumerate,float,indentfirst}
\usepackage{color}
\usepackage{comment}
\usepackage{ dsfont }
\usepackage{ mathdots }
\def\be{\begin{eqnarray}}
\def\ee{\end{eqnarray}}
\def\nn{\nonumber}

\def\Tr{{\rm Tr}\,}

\def\R{$\mathcal{R}$}
\def\r{\mathcal{R}}

\usepackage[]{breqn}
\usepackage{comment}
\def\K{\mathcal{K}}
\def\L{\mathcal{L}}
\def\A{\mathcal{A}}
\def\P{\mathcal{P}}
\def\H{\mathcal{H}}
\def\W{\mathcal{W}}
\def\D{\Delta}
\def\l{\lambda}
\def\m{\mu}
\def\n{\nu}
\def\a{\alpha}
\def\b{\beta}
\def\g{\gamma}

\def\nn{\nonumber}

\def\uqslN{$\mathcal{U}_q(sl_N)$}
\def\uqsltwo{$\mathcal{U}_q(sl_2)$}
\def\uqslthree{$\mathcal{U}_q(sl_3)$}
\def\uqslfour{$\mathcal{U}_q(sl_4)$}
\def\U{U}
\usepackage{lscape}

\usepackage{pinlabel}
\usepackage{graphicx}

\def\bc{\begin{center}}
	\def\ec{\end{center}}

\newcommand{\pic}[2]{\raisebox{-.45\height}{\includegraphics[scale=#2]{#1}}}

\newcommand{\pick}[2]{\raisebox{-.5\height}{\includegraphics[scale=#2]{#1}}}

\def\trefoilzero{\pic{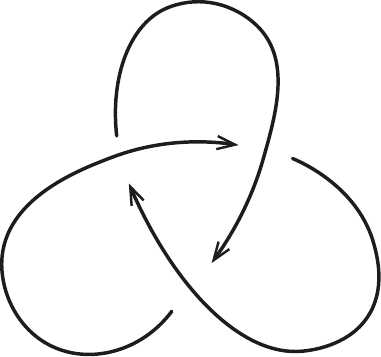}{.550}}
\def\trefoilB{\pic{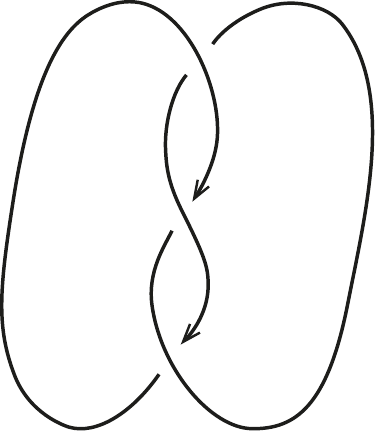}{.550}}
\def\trefoilBW{\pic{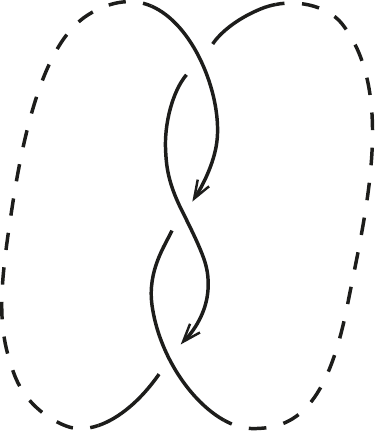}{.550}}
\def\trefoiljust{\pic{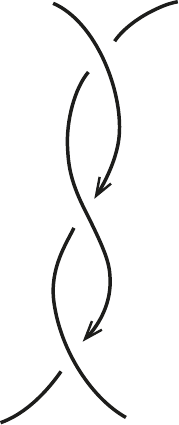}{.550}}
\def\trefoilBWcut{\pic{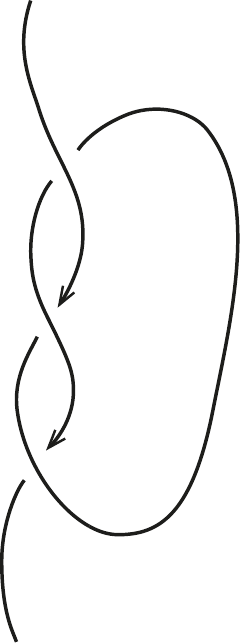}{.550}}

\def\eight{\pic{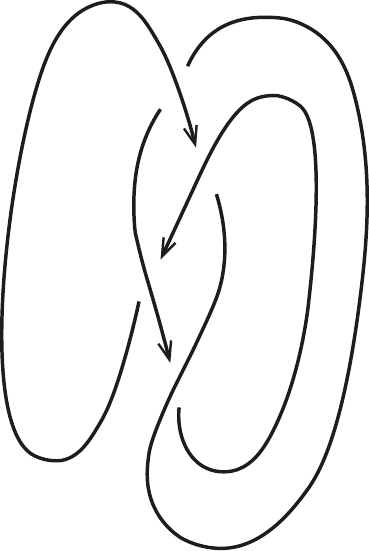}{.700}}
\def\five{\pic{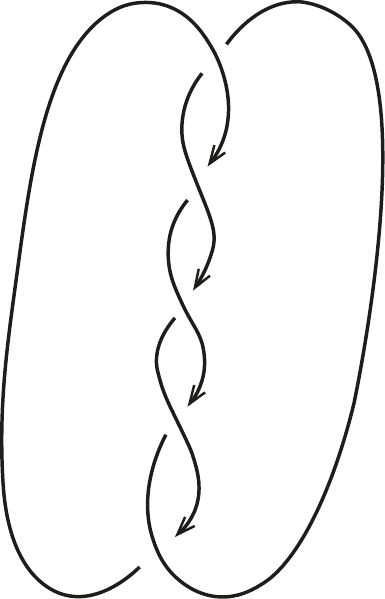}{.700}}
\def\hopf{\pic{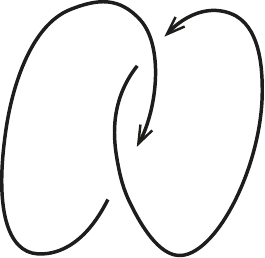}{.700}}
\def\trefoil{\pic{Pictures/trefoilBraid}{.700}}

\def\eightcut{\pic{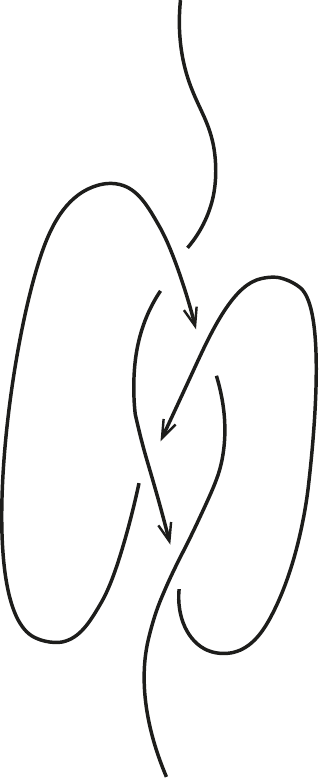}{.700}}
\def\fivecut{\pic{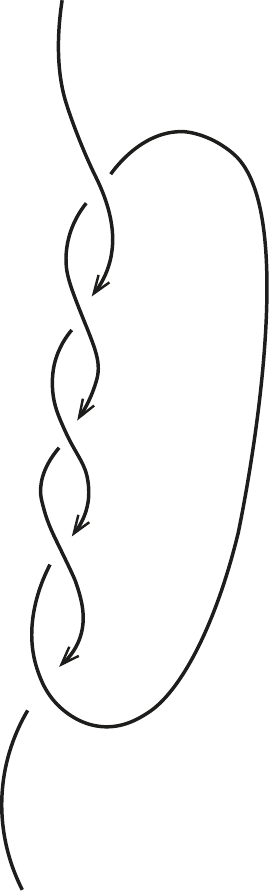}{.700}}
\def\hopfcut{\pic{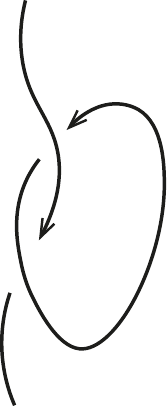}{.700}}
\def\trefoilcut{\pic{Pictures/trefoilBcut}{.700}}

\def\whlink{\pic{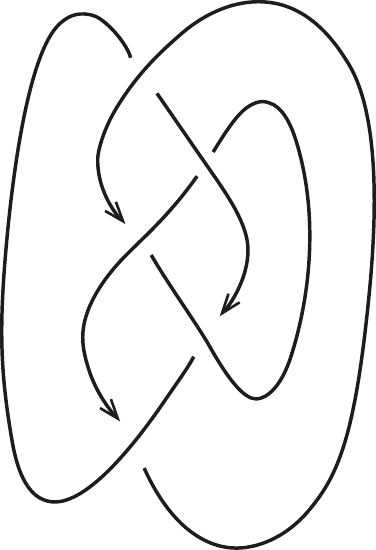}{.700}}
\def\brlink{\pic{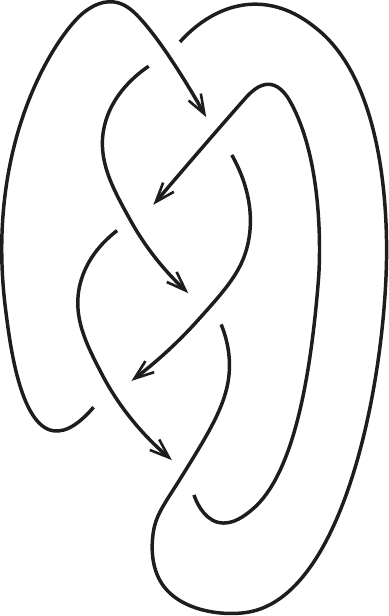}{.700}}
\def\lseven{\pic{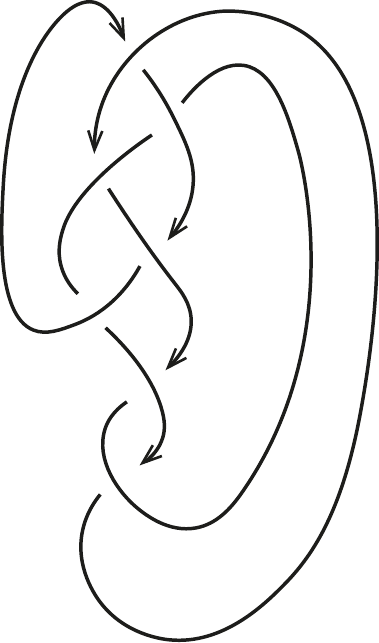}{.700}}

\def\whlinkcut{\pic{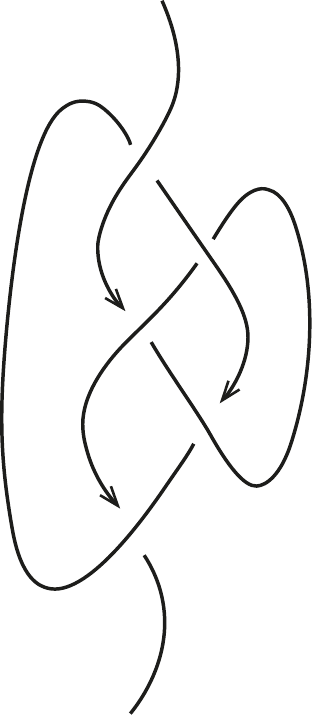}{.700}}
\def\brlinkcut{\pic{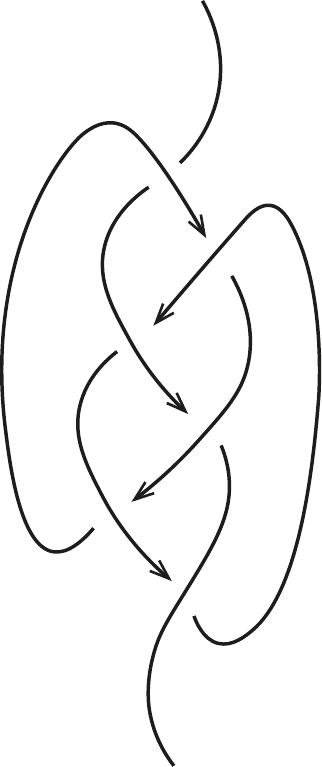}{.700}}
\def\lsevencut{\pic{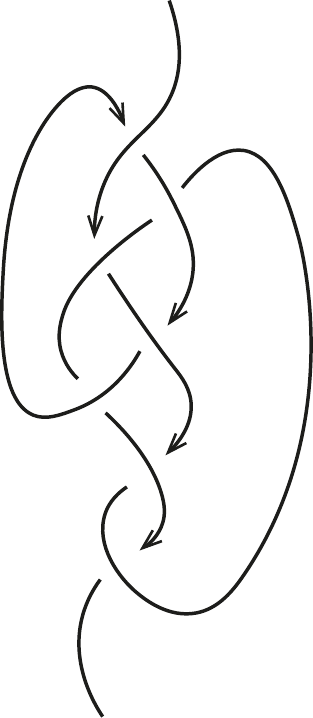}{.700}}

\def\ket{\pick{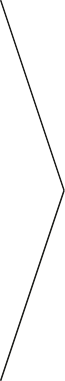}{1.0}}
\def\miniket{\pick{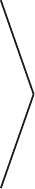}{1.0}}

\definecolor{red}{rgb}{1,0,0}
\definecolor{orange}{rgb}{1,0.5,0}
\definecolor{violet}{rgb}{0.7,0,1}

\hoffset= - 1.0in         

\begin{document}
\title{\vspace{1cm}{\Large {\bf Invariants of knots and links at roots of unity  }\vspace{.2cm}}}
	\author{\ {\bf Liudmila Bishler$^{a,b}$}, \ {\bf Andrei Mironov$^{a,b,c}$}, \ {\bf Andrey Morozov$^{b,c,d}$}
	\date{ }
}

\maketitle

\vspace{-5.5cm}

\begin{center}
\hfill FIAN/TH-04/22

	\hfill MITP/TH-11/22
	
	\hfill ITEP/TH-14/22
	
	\hfill IITP/TH-12/22
\end{center}

\vspace{3.0cm}

\begin{center}
	
	$^a$ {\small {\it Lebedev Physics Institute, Moscow 119991, Russia}}\\
	$^b$ {\small {\it Kurchatov Institute, Moscow, 123182, Russia}}\\
	$^c$ {\small {\it Institute for Information Transmission Problems, Moscow 127994, Russia}}\\
	$^d$ {\small {\it Moscow Institute of Physics and Technology, Dolgoprudny 141701, Russia }}
\end{center}

\vspace{1cm}

\centerline{ABSTRACT}

\bigskip

{\footnotesize
We present a comprehensive classification of invariants of knots and links associated with irreducible representations of \uqslN{}, when the parameter of quantization $q$ is a root of unity. We demonstrate that, besides the standard HOMFLY-PT invariants, which are associated with representations with highest and lowest weights, non-trivial invariants can be associated only with nilpotent representations with parameters. We define the corresponding invariants and discuss their relations with standard invariants at particular values of parameters.
}

\bigskip

\bigskip

\section{Introduction}

Invariants of knots and links that correspond to representations of $\mathcal{U}_q(sl_N)$ turn out to emerge in various branches of theoretical physics and nowadays attract a lot of attention. In particular, these are natural objects in Chern-Simons theory \cite{CS} (Wilson averages \cite{Witten}), in $2d$ conformal field theory \cite{CFT}: in the Wess–Zumino–Witten model (conformal blocks \cite{Kaul}) and in the minimal models (conformal blocks \cite{GMMM,Fabio}), etc. On the mathematical side, they are a part of theory of invariants of $3d$ manifolds. Of a special interest are the invariants at values of $q$ taking values at roots of unity, which just corresponds to Chern-Simons theory and others.

In particular, in Chern-Simons theory, the parameter $q$ is a function of coupling constant (level in the Wess–Zumino–Witten model) :
\begin{equation}
q=e^{\cfrac{2\pi i}{k+N}}.
\end{equation}
For Chern-Simons theory to be gauge invariant, this parameter $k$ should be integer \cite{WitCS}, thus, in this case, $q$ should be a root of unity. This means that understanding of how observables of Chern-Simons theory behave requires studying the structure of knot and link invariants for $q$ equal to a root of unity.

Some of invariants at $q$ taking values at roots of unity, ADO invariants were first introduced by Y. Akutsu, T. Deguchi and T. Ohtsuki in \cite{ADO}. On the physical side, the ADO invariants are related to other invariants, including the recent homological blocks \cite{GPV,GPPV,GM}. Moreover, it was proved that the full set of the ADO invariants of a knot is equivalent to a full set of the colored Jones invariants of that knot \cite{Wil,CGP,Gukov20,BDGG}.
Thus, in this paper, we are interested in invariants associated with representations of $\mathcal{U}_q(sl_N)$, when the parameter of quantization $q$ is a root of unity \cite{milaroots2}.

In this paper, we consider $q$ being the $2m$-th primitive root of unity, however all the results immediately extend to odd degrees. 

In the case of indeterminate $q$ (when $q$ is not a root of unity), the invariants associated with $\mathcal{U}_q(sl_N)$ are the HOMFLY-PT polynomials \cite{HOMFLY-PT} and their specializations: the Jones \cite{Jones} and the Alexander \cite{Alex} polynomials. The colored HOMFLY-PT polynomials can be calculated for various irreducible finite-dimensional representations of $\mathcal{U}_q(sl_N)$, for instance, with the Reshetikhin-Turaev (RT) method \cite{turaev} and its modern versions \cite{RTmod}.

Let us note that the quantized universal enveloping algebra of $sl_N$ (\uqslN{}) is generated by elements $E_i$, $F_i$, $K_i$, $K_i^{-1}$ $(i = 1, \dots, N-1)$, which satisfy the relations:
\begin{equation}
    \begin{array}{lll}
             K_i E_j = q^{a_{ij}}E_jK_i, \,\,\,\,\,\,&
    K_i F_j = q^{-a_{ij}}F_jK_i, \,\,\,\,\,\,&
    [E_i,F_j] = \delta_{ij}{K_i-K_i^{-1} \over q-q^{-1}},
    \end{array}
\end{equation}
\begin{equation*}
    \begin{array}{lll}
         [K_i,K_j] = 0, \,\,\,\,\,\, &
    [E_i,E_j] = 0 \,\,\,\,\,\, \text{for}\,\,\,\,\,\, |i-j|>1, &
     [F_i,F_j] = 0 \,\,\,\,\,\, \text{for}\,\,\,\,\,\, |i-j|>1 \\
    \end{array}
\end{equation*}

\begin{equation}
    \begin{array}{l}
        E_i^2 E_{i\pm 1} - (q+q^{-1})E_i E_{i\pm 1}E_i+ E_{i \pm 1}E_i^2 = 0,\\
         F_i^2 F_{i\pm 1} - (q+q^{-1})F_i F_{i\pm 1}F_i+ F_{i \pm 1}F_i^2 = 0,\\
    \end{array}
\end{equation}
where $(a_{ij})_{i,j=1,\dots,N-1}$ is the Cartan matrix of $sl_N$: $a_{i,i}=2$, $a_{i,i+1}=-1$, ${a_{i,j}}=0$ for $|i-j|>1$.

When $|q|<1$, the representation structure of $\mathcal{U}_q(sl_N)$ is the same as that of the non-deformed algebra $\mathcal{U}(sl_N)$. When $q$ is a root of unity, the representation structure of $\mathcal{U}_q(sl_N)$ changes \cite{Klimyc, Arnaudon}, and we will call such an algebra $\U_q(sl_N)$. Operators $E^m$, $F^m$ and $K^m$ belong to the center of $\U_q(sl_N)$ which restricts dimensions of irreducible representations. Among irreducible finite-dimensional representations of $\U_q(sl_N)$, there are both ordinary representations with the highest and the lowest weights $L_{m,N}$, which coincide with representations of $\mathcal{U}_q(sl_N)$ and have dimensions less than $m^{N(N-1)/2}$, and new types of representations of dimensions $m^{N(N-1)/2}$ with parameters: cyclic $\mathcal{U}_{m,N}$ and semi-cyclic $V_{m,N}$ ones, and nilpotent representations with parameters $W_{m,N}$.
These four types of representations produce invariants of knots and links, which can be calculated with the Reshetikhin-Turaev approach using $\mathcal{R}$-matrices, however the method should be modified. The cyclic and semi-cyclic representations give rise to trivial invariants of knots and links, and the invariants associated with representations $L_{m,N}$ coincide with the HOMFLY-PT polynomials. The most interesting invariants $\mathcal{P}_{m,N}(\lambda_i)$ are associated with the nilpotent representations with parameters $W_{m,N}(\lambda_i)$.

\begin{figure}[h]
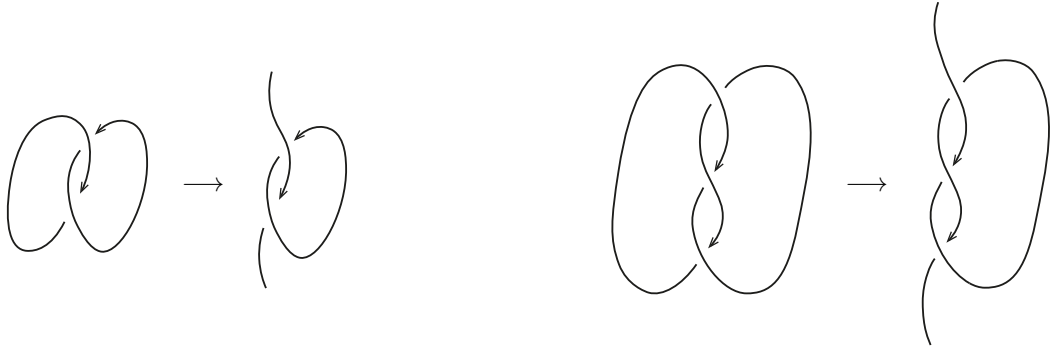

\begin{center}
$\quad \quad \quad \quad \quad
{\hopf \quad \longrightarrow \quad
\hopfcut}
\quad \quad \quad \quad \quad \quad \quad \quad \quad \quad
{\trefoil \quad \longrightarrow \quad
\trefoilcut}$
\end{center}
\caption{$(1,1)$-tangles corresponding to the Hopf link and  trefoil knot}
\label{hopftrefoilcut}
\end{figure}

A problem with the standard version of the RT method is that it produces invariants that are all equal to zero at roots of unity. It happens because the normalization coefficient (invariant of the unknot) is equal to zero in this case. A modification of the RT method enables us to define normalized (reduced) polynomials. The idea is the following: one should cut one strand of a knot/link and evaluate a polynomial of the corresponding $(1,1)$-tangle instead (see Fig.\ref{hopftrefoilcut}). It is possible due to existing correspondence between knots/links and (1,1)-tangles \cite{tangles}. The procedure in this case produces reduced polynomials, which specifically depend on the color of the line that has been cut. In order to restore the symmetricity of its components and get an invariant, one should divide the polynomials by a normalization coefficient $\Xi_{m,N}(\l_i^{(1)})$. This coefficient was known for $\mathcal{U}_q(sl_2)$ \cite{ADO}. In this paper, we first found it $\mathcal{U}_q(sl_3)$ and $\mathcal{U}_q(sl_4)$ and then generalized for $\U_q(sl_N)$ at arbitrary $N$.

The idea to use tangles instead of knots (which is equivalent to using the modified quantum trace operation) was implemented by Y. Akutsu, T. Deguchi and T. Ohtsuki in \cite{ADO}, where they also calculated normalization coefficient for $\mathcal{U}_q(sl_2)$ that restores invariance of the answers. M. Rosso mentioned the use of modified quantum trace operation in \cite{Rosso}.  J. Murakami in his work \cite{Murakami} used the $\mathcal{R}$-matrix approach (based on the universal $\mathcal{R}$-matrix) and tangles to define the invariants of knots/links for $\mathcal{U}_q(sl_2)$ at roots of unity. In this paper, we extend their results, considering the non-nilpotent representations of $\U_q(sl_N)$, and present a comprehensive study of invariants of knots and links corresponding to $\mathcal{U}_q(sl_N)$ at roots of unity.

\bigskip
\bigskip

The structure of the paper is as follows. First of all, we present our main results in section \ref{secRes}: investigation of $\mathcal{R}$-matrices, corresponding to irreducible finite-dimensional representations of $\mathcal{U}_q(sl_N)$ at roots of unity, definition of invariants and comparison of these invariants with the HOMFLY-PT and Alexander polynomials. In section \ref{secreps}, we present the representation structure of $\mathcal{U}_q(sl_N)$ at roots of unity and discuss the $\mathcal{R}$-matrices. In section \ref{secRT}, we give an overview of the Reshetikhin-Turaev method of evaluating knot/link invariants and discuss how this method is modified to calculate the invariants at roots of unity. In section \ref{secSl2}, we discuss $\U_q(sl_2)$ case: we give explicit formulas for representations, the $\mathcal{R}$-matrices, define invariants and compare them with the Jones and Alexander polynomials. In section \ref{secSl34}, we present the same consideration for $\U_q(sl_3)$ and $\U_q(sl_4)$. In Appendices A and B, we list the invariants associated with the nilpotent representations, we also list the Alexander polynomials that we use in this paper in Appendix C.





\subsection*{Notation}
In this paper, we work with even degrees of roots of unity, $q^{2m} = 1$. In the case of odd degrees, the representations are the same, however the answers for invariants of links and knots slightly differ. We denote $\mathcal{U}_q(sl_N)$ when $q$ is a root of unity as $\U_q(sl_N)$.

In explicit formulas for particular knot/links, $q$ is a primitive root of unity of the degree $2m$  (i.e. there is no $k<2m$ such that $q^k=1$). Primitive roots correspond to roots of the cyclotomic polynomials $C_{2m}$, which we used for simplification of expressions.
Here is the list of the ones that we used:

\begin{equation}
    \begin{array}{ll}
        C_4 = q^2+1, &
        C_6 = q^2-q+1,\\
        C_8 = q^4+1, &
        C_{10} = q^4-q^3+q^2-q+1, \\
        C_{12} = q^4-q^2+1,\,\,\,\,\,\,\,\, &
        C_{14} = q^6-q^5+q^4-q^3+q^2-q+1.\\
    \end{array}
\end{equation}

We work with $\mathcal{U}_q(sl_N)$, and everywhere $N-1$ means the rank of $sl_N$. We discuss irreducible finite-dimensional representations $T_{m,N}(\lambda_i)$ of $\U_q(sl_N)$, when $q$ is a root of unity. Representations depend on algebra, half-degree of root $m$ and the value of operators $E^m$, $F^m$ and $K^m$, which enter the parameters $\lambda_i$. We mention the following types of representations of $\U_q(sl_N)$ at roots of unity:
\begin{itemize}
    \item $L_{m,N}$ --- the highest and the lowest weight representations without parameters,
    \item $U_{m,N}$ --- cyclic representations,
    \item $V_{m,N}$ --- semi-cyclic representations,
    \item $W_{m,N}$ ---  the highest and the lowest representations with parameters (nilpotent representations with parameters).
    \end{itemize}

We mention the following types of polynomial invariants of knots/links $\L$ (we use $\L$ to denote both knots and links):
\begin{itemize}
\item $H^{\L}_R (A = q^N,q) $ are the reduced HOMFLY-PT polynomials  (divided with the HOMFLY-PT polynomial of unknot in representation $R$) associated with representation $R$ of $\mathcal{U}_q(sl_N)$, when $q$ is not a root of unity. The HOMFLY-PT polynomials in various representations of $\mathcal{U}_q(sl_N)$ can be found in \cite{knotebook}

\item $\A^{\K}(q)$ are the Alexander polynomials of knots, which coincide with the specialization of the HOMFLY-PT polynomials in fundamental representation at $A=1$: $\A^{\K}(q) = H_{\square}^{\K}(A=1, q)$, they can be found in the Rolfsen table of knots ($q^2=t$) \cite{katlas} and in Appendix C.

\item $\Delta^{\L}(\{q^{r_i}\})$ are the Alexander polynomials of links, which are also connected with the HOMFLY-PT polynomials in symmetric representations at $A=1$: $\Delta^{\L}(q) = \left. {H_{\{[r_i]\}}^{\L}(A, q) \over 1-A^2} \right|_{A\rightarrow 1}$, they can be found in the Rolfsen table of knots \cite{katlas} and in Appendix C.

\item  We denote $P_{m,N}^{\L}$ {\bf the polynomials} associated with the nilpotent representations with parameters of $\mathcal{U}_q(sl_N)$, when $q$ is the $2m$-th primitive root of unity. The polynomials $P_{m,N}^{\K}$ of knots are listed in Appendix A.

\item We denote $\P_{m,N}^{\L}$ {\bf the invariants} of knots/links associated with the nilpotent representations with parameters of $\mathcal{U}_q(sl_N)$, when $q$ is the $2m$-th primitive root of unity . They are proportional to  $P_{m,N}^{\L}$. We list invariants of links $\P_{m,N}^{\L}$ in Appendix B.
\end{itemize}

The polynomials $P_{m,N}^{\L}$ and invariants $\P_{m,N}^{\L}$ depend on parameters of the nilpotent representations $W_{m,N}$:

\begin{itemize}
    \item\uqsltwo{}: $P_{m,2}^{\L} (\l)$ for knots and same-colored links, $P_{m,2}^{\L}(\l^{(1)},\l^{(2)},\dots)$ for links with different colors on different components,
    \item \uqslthree{}:
    $P^{\L}_{m,3} (\l_1,\l_2)$ for knots and same-colored links, $P^{\L}_{m,3}(\l^{(1)}_1,\l^{(1)}_2,\l^{(2)}_1,\l_2^{(2)}\dots)$ for links with different colors on different components,
     \item \uqslfour{}:
    $P^{\L}_{m,4} (\l_1,\l_2,\l_3)$ for knots and same-colored links, $P^{\L}_{m,4}(\l^{(1)}_1,\l^{(1)}_2,\l^{(1)}_3,\l^{(2)}_1,\l_2^{(2)},\l_3^{(2)},\dots)$ for links with different colors on different components,
\end{itemize}

If we color different $l$ components of a link with colors $\l_{i}^{(k)}$, where $1\leq k \leq l$, $1\leq i \leq N-1$ and $\l_i^{(1)}$ is the color of the open component on the link diagram, then the invariant corresponding to the representations $W_{m,N}(\lambda_i^{(k)})$ is the following

\begin{equation}
    \P_{m,N}^{\L} (\l_i^{(k)}) = {P_{m,N}^{\L}(\l_i^{(k)}) \over \Xi_{m,N}(\l^{(1)}_i)}.
\end{equation}
Determination of the coefficient $\Xi_{m,N}(\l^{(1)}_i)$ is one of the main results of this paper.

\bigskip
We also use the quantum numbers: $[x] = {q^{x}-q^{-x} \over q-q^{-1}}$, the bracket  $\{x\} = x-x^{-1}$, and the quantum Pochhammer symbol $[x;n] = \prod_{i=0}^{n-1}[x-i]$.

\section{Main results \label{secRes}}

In this paper, we define invariants of knots and links corresponding to irreducible finite dimensional representations of $\mathcal{U}_q(sl_N)$, when the parameter of quantization $q$ is a root of unity. We study all possible representations of this kind, and, hence, complete the list of all possible knot/link invariants described by $\mathcal{U}_q(sl_N)$.

We constructed representations, calculated \R-matrices and, using the modified Reshetikhin-Turaev method, evaluated invariants of knots and links. It allowed us to find the generic normalization coefficient, and to compare the resulting invariants with HOMFLY-PT and Alexander. We list our main results below.

\subsection*{\R-matrices}
Irreducible finite dimensional representations of $\U_q(sl_N)$ were classified and constructed in \cite{Arnaudon}. Four different types of representations give us the following \R-matrices at $q$ such that $q^{2m}=1$:
\begin{itemize}
    \item For representations with the highest and the lowest weights $L_{m,N}$, the \R-matrices coincide with not-a-root-of-unity case, hence, we do not consider them here.
    \item The cyclic representations $U_{m,N}$ and semi-cyclic representations $V_{m,N}$ produce the \R-matrices that give rise to trivial invariants of knots and links.
    \item Nilpotent representations with parameters $W_{m,N}(\lambda_i)$ produce the \R-matrices $\r_{m,N}(\l_i)$ that allow one to get non-trivial polynomials of knots and links. This is the main point of our interest in this paper. We manifestly evaluated $\r_{m,2}$, $\r_{2,3}$, $\r_{3,3}$, $\r_{4,3}$ and $\r_{2,4}$.

\end{itemize}

\subsection*{Definition of invariants}

One of the main results of this paper is the definition of invariants $\P^{\L}_{m,N}(\lambda)$ of knots and links associated with nilpotent representations $W_{m,N}(\l^{(k)}_i)$ of $\mathcal{U}_q(sl_N)$ for arbitrary $N$, when $q$ is a root of unity.

In order to calculate an invariant, one should color components of a link (that consists of $j$ components) with representations $W_{m,N}(\l^{(k)}_i)$ ($\l^{(k)}_i = \{\l_1^{(k)},\dots,\l_{N-1}^{(k)}\}$, $1\leq k \leq j$), cut one strand of the link and apply the RT method to the $(1,1)$-tangle. This procedure gives rise to a reduced polynomial $P^{\L}_{m,N}(\l)$, which depends on the color of an open component.
\begin{figure}[H]
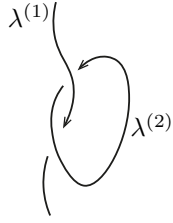

\bc
{\labellist
 \pinlabel{$\l^{(1)}$} at -7 110
 \pinlabel{$\l^{(2)}$} at 60 50
\endlabellist
\hopfcut}
\ec
\caption{Colored $(1,1)$-tangle corresponding to the Hopf link}
\label{coloredhopf}
\end{figure}
The very procedure has been known earlier \cite{ADO,Murakami,Rosso}, but the subtle point is to fix a proper normalization coefficient. Let the color of the open component be $\l^{(1)}_i$, then we have found that the proper normalization coefficient, which, in particular, provides a link invariant which is symmetric with respect to all link components is the following:

\begin{equation}
\boxed{
   \Xi_{m,N}(\l^{(1)}_i) =  \prod_{\alpha \in \Phi^{+}_N} \xi_m( \l^{(1)}_\alpha q^{|\alpha|}),
   } \label{norm}
\end{equation}
where $\alpha$ are positive roots $\Phi^{+}_N$ of $sl_N$,  $\alpha=\sum_{k=i}^j\alpha_k$ ($i\leq j < N $), where $\alpha_k$ are simple roots of $sl_N$, $|\alpha|=j-i$,
\begin{equation}
    \xi_m(\lambda)=\prod_{i=0}^{m-2} \{\lambda q^{-i}\}
\end{equation}

\begin{equation}
    \begin{array}{l}
         \Xi_{m,2}(\l) = \xi_m(\l),\\
\Xi_{m,3}(\l) = \xi_m(\lambda_1)\, \xi_m(\l_2)\, \xi_m(q\lambda_1 \l_2),\\
\Xi_{m,4}(\l) = \xi_m(\lambda_1)\, \xi_m(\l_2)\, \xi_m(\l_3)\,\xi_m(q\lambda_1\l_2)\,\xi_m(q\l_2\l_3)\,\xi_m(q^2\lambda_1\l_2\l_3)
    \end{array}
\end{equation}
Hence, (unreduced) invariants of knots and links are
\begin{equation}
\boxed{
    \P_{m,N}^{\L}(\l^{(k)}_i) = {P_{m,N}^{\L}(\l_{i}^{(k)}) \over \Xi_{m,N}(\l^{(1)}_i)}.}
\end{equation}
Although $P_{m,N}^{\L}$ is a (Laurent) polynomial of $\lambda$'s, the invariant $\P_{m,N}^{\L}$ is a rational function.
There is also an issue of framing, we discuss it separately in sec.\ref{frame}.

\subsection*{Connection with HOMFLY-PT}

The parameters of representations $\lambda_1, \, \l_2, \dots$ in our calculations play the role of the highest weights  ($K_i v_0 = \l_i v_0 $), and there is a specific correspondence between the HOMFLY-PT polynomials $\mathcal{H}_{T}^{\L}(A=q^N,q)$ in representation $T$ corresponding to the Young diagram $[(N-1)(m-1), (N-2)(m-1), \dots, (m-1)] $ and $P_{m,N}^{\L}$ at special values of these representation parameters:

\begin{equation}
    \boxed{
 \left.   P_{m,N}^{\mathcal{L}}\,(q,\lambda_i = q^{m-1})\right|_{q^{2m}=1} = \left. H^{\L}_{[(N-1)(m-1), (N-2)(m-1), \dots, (m-1)]} (A=q^N,q)\right|_{q^{2m}=1}
    }
\end{equation}
This connection exists because the nilpotent representation $W_{m,N}$ is a generalization of the  representation  $[(N-1)(m-1), (N-2)(m-1), \dots, (m-1)] $. In representation $W_{m,N}$, the highest weights $\l_i$ are not fixed, and when they are equal to the highest weights of $[(N-1)(m-1), (N-2)(m-1), \dots, (m-1)] $, the corresponding polynomials coincide with the HOMFLY-PT polynomials.

\subsection*{Connection with Alexander polynomials}
We also uncovered some interesting connections between $P_{m,N}^{\L}$ and the Alexander polynomials in particular cases.

First of all, the Alexander polynomials of knots $\A^{\mathcal{K}}(q)$ are connected with the HOMFLY-PT polynomials in fundamental representation at $A=1$ specialization: $\A(q) = H_{[1]}^{\mathcal{K}}(A=1,q)$, and, as it was discovered in \cite{Rosso}, coincide with polynomials $P^{\L}_{2,2}$ corresponding to the nilpotent representation with parameter $W_{2,2}(\l)$ of  \uqsltwo{} when $q^4=1$:
\begin{equation}
    P^{\K}_{2,2}\,(\lambda) = \A^{\K} (\lambda).
\end{equation}
This connection can be extended to links: we found that the multivariable Alexander polynomials of links $\D^{\L}(u,v,\dots)$ coincide with invariants $\P^{\L}_{2,2}$ corresponding to the 4-th root of unity \uqsltwo{}, calculated for links with different colors on different components.

\begin{equation}
    \P^{\L}_{2,2}(\lambda^{(1)}, \l^{(2)}, \dots) = \D^{\L}(\lambda^{(1)},\l^{(2)},\dots)
\end{equation}

Besides these, very general relations, we found, in the case of knots, connections between polynomials $P^{\K}_{2,3}$ and $P^{\K}_{3,3}$ for the 4-th and 6-th roots of unity of \uqslthree{} and the Alexander polynomials:
\begin{equation}
     P^{\K}_{2,3}\,(\lambda_1, \l_2=1) = \A^{\K} (\lambda^{2}_1),
\end{equation}
\begin{equation}
     P^{\K}_{3,3}\,(q,\lambda_1, \l_2=1) =\A^{\K} (\lambda_1) \A^{\K} (\lambda^{3}_1),
\end{equation}
These expressions are the same for $\lambda_1 = 1$ because $P_{m,3}^{\K}\,(\lambda_1,\l_2) = P_{m,3}^{\K}\,(\l_2,\lambda_1)$.
In the case of \,\uqslfour{}, the symmetry between the parameters is more complicated, and there are the following connections between $ P_{2,4}^{\K}(q,\l_1,\l_2,\l_3)$ and the Alexander polynomials.
\begin{equation}
    P_{2,4}^{\K}(q,\l_1=1,\l_2,\l_3=1) = \A^{\K}(\l_2^2),
\end{equation}
\begin{equation}
P_{2,4}^{\K}(q,\l_1=1,\l_2=1,\l_3)  = \A^{\K}(\l_3)\A(\l_3^2),
\end{equation}
\begin{equation}
    P_{2,4}^{\K}(q,\l_1,\l_2=1,\l_3=1)  = \A^{\K}(\l_1)\A(\l_1^2).
\end{equation}

Invariants of links $\P_{2,3}^{\L}$ are also connected with the Alexander polynomials of links $\D^{\L}$, however the connection is more elaborate:
\begin{equation}
    \P_{2,3}^{L_2 a_1} (q,\l_1^{(1)},\l_2^{(1)},\l_1^{(2)},\l_2^{(2)}) = q \, \D^{L_2a_1},
\end{equation}
\begin{equation}
    \P_{2,3}^{L_5 a_1}(q,\l_1^{(1)},\l_2^{(1)} = q,\l_1^{(2)},\l_2^{(2)} = q) = -4q \left(\left(\D^{L_5 a_1}(\l_1^{(1)},\l_1^{(2)} )\right)^2-1\right),\\
\end{equation}
\begin{equation}
    \P_{2,3}^{L_6 a_4}(q, \l_1^{(1)},\l_2^{(1)} = q,\l_1^{(2)},\l_2^{(2)} = q,\l_1^{(2)},\l_2^{(2)} = q) = 8q \left(\left(\D^{L_6 a_4}(\l_1^{(1)},\l_1^{(2)},\l_1^{(3)} )\right)^2-2\right).\\
\end{equation}

\section{Irreducible representations of $\U_q(sl_N)$\label{secreps}}
In this section, we describe the representation structure of $\U_q(sl_N)$, when the deformation parameter $q$ is $2m$-th root of unity.
This section is mostly follows the work by B.~Abdesselam, D.~Arnaudon and A.~Chakrabarty \cite{Arnaudon}.

\subsection{General structure of representations}
Finite dimensional irreducible representations of $\U_q(sl_N)$ include the usual representations with the highest and the lowest weights as well as new types of representations, which do not exist in the not-a-root-of-unity case: cyclic (periodic), semi-cyclic (semi-periodic) and representations with the highest and the lowest weights with parameters (we call them nilpotent representations with parameters). There are also exotic types of representations: atypical and partially periodic, 
we do not consider them in this paper. The dimensions of all irreducible representations do not exceed $m^{N(N-1)/2}$. We are especially interested in cyclic, semi-cyclic and nilpotent representations with parameters, they all have dimension $m^{N(N-1)/2}$, which depends on the degree of the root.

These new types of representations exist because in the root-of-unity case the center of algebra consists not only of the Casimir operators $\mathcal{C}_i$, but also includes the operators $E_{\alpha}^m$, $F_{\alpha}^m$, $K_i^m$, where $\alpha$ are the positive roots. These operators act as scalars on elements of the algebra and provide $N^2-1$ continuous parameters. These parameters are indirectly related with continuous parameters in the formulas of generators from sec.\ref{gens} $c_{jl}$ and $p_{jl}$.

The list of representations for each particular root degree can be found in the Table, $q^{2m}=1$.

\begin{table}[H]
    \centering
    \small
    \begin{tabular}{|c|c|c|c|}
    \hline
    &&&\\
         dimension & name & $E^m$, $F^m$ & description \\
         &&&\\
         \hline
         &&&\\
         $d<m^{N(N-1)/2}$ & $L_{m,N}$ & $E_{\alpha}^m = 0$, & the highest and the lowest weight representations\\ &&$F_{\alpha}^m = 0$&  \\
         &&& \\
         \hline
         &&&\\
         $d=m^{N(N-1)/2}$ &  $U_{m,N}(p_{jl},c_{jl})$ & $E_{\alpha}^m \neq 0$,  & \textbf{cyclic} representation\\
        &&$F_{\alpha}^m \neq 0$& without the highest and the lowest weights \\
         &&&\\
         \cline{2-4}
         &&&\\
        &  $V_{m,N}(p_{iN},c_{jl})$   &$E_{\alpha}^m = 0$ or& \textbf{semi-cyclic} representation\\
         & & $F_{\alpha}^m = 0$ & with the highest or the lowest weight \\
         &&&\\
         \cline{2-4}
         &&&\\
         & $W_{m,N}(p_{iN})$ & $E_{\alpha}^m = 0$, & the highest and the lowest weight (\textbf{nilpotent}) \\
         &&$F_{\alpha}^m = 0$& representation with parameters \\
         &&&\\
         \hline
    \end{tabular}
    \caption{Irreducible representations of $\U_q(sl_N)$ at roots of unity}
    \label{tab:my_label}
\end{table}

The number of continuous parameters for each representation is the following:
\begin{itemize}
    \item $L_{m, N}$ does not contain continuous parameters.
    \item  $U_{m,N}(p_{jl},c_{jl})$ contains the biggest set of parameters: $N^2-1$ in total, $c_{jl}$ for $1\leq j\leq l <N$ and $p_{j,l}$ for $1\leq j\leq l\leq N$.
    \item $V_{m,N}(p_{iN},c_{jl})$ : $(N-1)(N-2)/2$ in total, where there are $N-1$ of  $p_{iN}$s and $N(N-1)/2$ of $c_{jl}$s.
    \item $W_{m,N}(p_{iN})$ contains $N-1$ parameters $p_{iN}$.
\end{itemize}

\subsection{Action of the generators \label{gens}}
Each vector of representation is associated with a state $\left | p \right \rangle $ given by a set of parameters $p_{il}$, which defines actions of the generators on each particular vector

\begin{equation}
\left | p \right \rangle \,\,\,= \,\,\,
\left | \,\,
    \begin{array}{ccccccc}
    p_{1,N} & & p_{2,N} & \dots & p_{N-1,N} && p_{N,N} \\
    &&&&&\\
    & p_{1,N-1} & & \dots & & p_{N-1,N-1} & \\
    &&&&&&\\
    && \ddots & \dots & \iddots & \\
    &&&&&&\\
    && p_{12} && p_{22} & \\
    &&&&&&\\
    &&& p_{11} && \\
    \end{array}
    \right. \ket
\end{equation}

The generators $K_l^{\pm 1}$, $E_l$, $F_l$ act as follows
\begin{align}
K_l^{\pm 1} |p\rangle & =
 q^{\pm\left( 2\sum\limits_{i=1}^{l}p_{il}
              -\sum\limits_{i=1}^{l+1}p_{i,l+1}
              - \sum\limits_{i=1}^{l-1}p_{i,l-1} - 1 \right)} |p\rangle, \\
F_l |p\rangle & = \sum_{j=1}^l
 c_{jl} {P'_1(j,l;p) P'_2(j,l;p) \over P'_3(j,l;p) }
 |p_{jl}-1 \rangle,\\
 E_l |p\rangle & = \sum_{j=1}^l
 c_{jl}^{-1} {P''_1(j,l;p_{jl}+1) P''_2(j,l;p_{jl}+1)
                               \over P''_3(j,l;p_{jl}+1) }
 |p_{jl}+1 \rangle,
\end{align}

where the state $ |p_{jl}\pm 1 \rangle$ differs from $|p\rangle$ by the change of one element: $p_{jl} \rightarrow p_{jl}\pm 1$, and
{ \small
\begin{equation}
    \begin{array}{ll}
       P'_1(j,l;p)  = \prod_{i=1}^{l+1}               [\varepsilon_{ij}(p_{i,l+1}-p_{j,l}+1)]^{1-\eta_{ijl}}\,\,\,\,\,   & P''_1(j,l;p_{jl}+1)  = \prod_{i=1}^{l+1}   [\varepsilon_{ij}(p_{i,l+1}-p_{j,l})]^{\eta_{ijl}}\\
       P'_2(j,l;p)  = \prod_{i=1}^{l-1}             [\varepsilon_{ji}(p_{j,l}-p_{i,l-1})]^{\eta_{j,i,l-1}}\,\,\,\,\,   &  P'_2(j,l;p)  = \prod_{i=1}^{l-1}  [\varepsilon_{ji}(p_{j,l}-p_{i,l-1})]^{\eta_{j,i,l-1}} \\
       P'_3(j,l;p)  = \prod_{{i=1 \atop i\ne j}}^{l}       [\varepsilon_{ij}(p_{i,l}-p_{j,l})]^{1/2}
             [\varepsilon_{ij}(p_{i,l}-p_{j,l}+1)]^{1/2} \,\,\,& P''_3(j,l;p_{jl}+1)  = \prod_{{i=1 \atop i\ne j}}^{l}
             [\varepsilon_{ij}(p_{i,l}-p_{j,l}-1)]^{1/2}
        [\varepsilon_{ij}(p_{i,l}-p_{j,l})]^{1/2},
    \end{array}
\end{equation}
}
where $\varepsilon_{ij} = 1$ for $i\leq j$ and $\varepsilon_{ij} =-1$ for $i> j$, $\eta_{ijl}$ are discrete parameters that break the symmetry between $E$ and $F$, they can be $0$, $1$ or $1/2$. These parameters are not included into the list of continuous parameters of representations.

We discuss particular representations that we build with these formulas in sections \ref{repsSl2}, \ref{repsSl3} and \ref{repsSl4}.

\subsection{$\mathcal{R}$-matrix}
$\mathcal{U}_q(sl_N)$ is quasitriangular and admits the universal \R-matrix, which can be used in the RT method to calculate invariants of knots and links. The universal \R-matrix  acts as an intertwiner on the tensor product of irreducible finite dimensional representations of $\mathcal{U}_q(sl_N)$, and is explicitly given by
\begin{equation}
\mathcal{R}_u = P \,q^{\sum_{i,j} a^{-1}_{i,j} h_i \otimes h_j} \overrightarrow{\prod_{\beta \in \Phi^{+}}} {\rm exp}_{q}     \left ( (q - q^{-1}) E_{\beta} \otimes F_{\beta}   \right),
\label{UniR}
\end{equation}
where $P (x \otimes y) = y \otimes x$, $\Phi^{+}$ are the positive roots, $q^{h_i} = K_i$, ${\rm exp}_q A = \sum_{m = 0}^{\infty} \frac{A^m}{[m]_q !} q^{m(m-1)/2}$.
It can also be applied to the nilpotent representations with parameters $W_{m,N}(p_{iN})$, and used for calculation of invariants of knots and links, when the parameter of quantization $q$ is equal to a root of unity.

We evaluated the $\mathcal{R}$-matrices for the nilpotent representations for an arbitrary $m$ for $\U_q(sl_2)$, for $m=2,3,4$ for $\U_q(sl_3)$, and for $m=2$ for $\U_q(sl_4)$, details of the calculations can be found in sections \ref{Rsl2}, \ref{Rsl3}, \ref{Rsl4}.

\subsubsection*{$\mathcal{R}$-matrix for cyclic and semi-cyclic representations}
In \cite{gomez}, it was shown that the intertwiner of the non-nilpotent representations (cyclic and semi-cyclic) exists. However, one cannot use it to calculate invariants of knots and links, because it satisfies the following conditions:
\begin{itemize}
    \item $\r(\xi, \xi) = 1$, which means that the intertwiner of identical representations is trivial and all invariants of knots that it produces, are also trivial.
    \item $\r(\xi_1,\xi_2) = \r^{-1}(\xi_2,\xi_1)$. This property makes this \R-matrix useless for calculation of link invariants, because the $\mathcal{R}$-matrix with such a property produces only trivial invariants \cite{RTT}.
\end{itemize}
It means that even though there exist non-trivial $\mathcal{R}$-matrices of the non-nilpotent representations (for example, eq.15 in \cite{gomez}), they do not produce non-trivial knot/link invariants.

\section{ Reshetikhin-Turaev method \label{secRT}}

In this section, we discuss basics of the RT approach to calculating knot/link invariants. The section mostly follows the paper \cite{MorSmir}.

We call two knots or links in $\mathds{R}^3$ ambient isotopy equivalent if they can be transformed into each other via smooth deformations in $\mathds{R}^3$. Invariants of ambient isotopy equivalent knots and links coincide. There are different methods to calculate polynomial invariants of knots, one of them is the Reshetikhin-Turaev method.


\subsection{Basics of RT method}
The group theoretical Reshetikhin-Turaev method \cite{turaev, MorSmir} is used to calculate HOMFLY-PT polynomials colored with different representations of $\mathcal{U}_q(sl_N)$. It can be applied with some alterations to calculation of polynomials corresponding to representations of $\U_q(sl_N)$ at roots of unity.

\begin{figure}[H]
  \centering
\begin{picture}(90,60)(150,0)

\qbezier(30,40)(15,30)(30,20)

\qbezier(30,40)(40,45)(50,40)
\qbezier(30,20)(45,15)(59,26)

\qbezier(50,40)(65,30)(70,20)
\qbezier(65,33)(67,35)(70,40)



\put(78,28){\mbox{$=$}}
\put(93,50){\line(0,-1){40}}


\qbezier(136,44)(134,46)(130,50)

\qbezier(136,16)(134,14)(130,10)
\qbezier(150,10)(120,30)(150,50)

\qbezier(142,20)(150,30)(142,40)



\put(163,28){\mbox{$=$}}

\put(185,50){\line(0,-1){40}}
\put(185,10){\line(0,1){25}}

\put(200,50){\line(0,-1){40}}
\put(200,10){\line(0,1){25}}


\put(250,10){\line(1,1){40}}
\put(250,50){\line(1,-1){17}}
\put(273,27){\line(1,-1){17}}

\put(257,53){\line(0,-1){7}}
\put(257,40){\line(0,-1){20}}
\put(257,14){\line(0,-1){7}}


\put(300,28){\mbox{$=$}}

\put(320,10){\line(1,1){40}}
\put(320,50){\line(1,-1){17}}
\put(343,27){\line(1,-1){17}}

\put(353,53){\line(0,-1){7}}
\put(353,40){\line(0,-1){20}}
\put(353,14){\line(0,-1){7}}

\end{picture}
  \caption{Reidemeister moves}\label{Reid}
\end{figure}
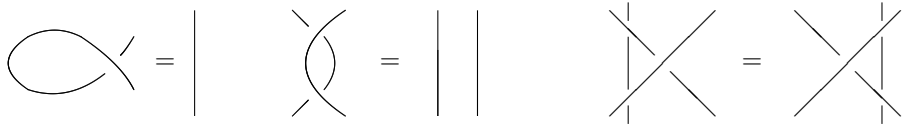

Within this approach, we use a projection of a knot on a plane with a fixed direction, which is called knot diagram. It is then clear which thread is above other in each crossing. The RT method is ambient isotopy equivalent because the operators that we use in this method preserve the Reidemeister moves (Fig.\ref{Reid}). According to the Reidemeister theorem, the two knots in  $\mathds{R}^3$ represented by their two-dimensional projections $P_1$ and $P_2$ are ambient isotopy equivalent if and only if $P_1$ can be deformed into $P_2$ via smooth deformations in the two-dimensional plane and using a finite set of Reidemeister moves.

\begin{figure}[h!]
    \centering
    \begin{picture}(300,40)(40,0)
    \put(0,5){\line(0,1){20}}
    \put(20,5){\line(0,1){20}}
    \qbezier(0,25)(10,40)(20,25)
    \put(35,12){\text{$=$}}
    \put(52,11){\text{$\mathcal{M}$}}
    \put(0,13){\vector(0,-1){2}}
    \put(20,13){\vector(0,1){2}}
     \put(100,5){\line(0,1){20}}
      \put(120,5){\line(0,1){20}}
    \qbezier(100,5)(110,-10)(120,5)
    \put(135,12){\text{$=$}}
    \put(152,11){\text{$\mathcal{M}$}}
    \put(100,13){\vector(0,-1){2}}
    \put(120,13){\vector(0,1){2}}
  \put(200,5){\line(0,1){20}}
      \put(220,5){\line(0,1){20}}
    \qbezier(200,25)(210,40)(220,25)
    \put(235,12){\text{$=$}}
    \put(252,11){\text{$\mathcal{M}^{-1}$}}
    \put(200,13){\vector(0,1){2}}
    \put(220,13){\vector(0,-1){2}}
      \put(300,5){\line(0,1){20}}
      \put(320,5){\line(0,1){20}}
    \qbezier(300,5)(310,-10)(320,5)
    \put(335,12){\text{$=$}}
    \put(352,11){\text{$\mathcal{M}^{-1}$}}
    \put(300,12){\vector(0,1){2}}
    \put(320,12){\vector(0,-1){2}}
    \end{picture}
     \caption{Four types of turning points}\label{turn}
\end{figure}
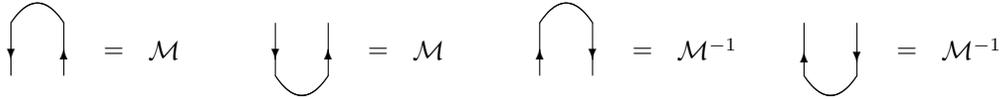

The knot diagram consists of crossings and turning points, which we associate with operators $\mathcal{R}$ and $\mathcal{M}$ correspondingly (Fig.\ref{crossings},\ref{turn}).
\begin{figure}[h!]
  \centering
\begin{picture}(90,50)(-70,0)
\put(50,10){\text{$\mathcal{R}^{-1}$}}
\put(34,10){\text{$=$}}

\put(0,0){\line(1,1){30}}
\put(0,30){\line(1,-1){13}}
\put(30,0){\line(-1,1){13}}

\put(5,5){\vector(-1,-1){2}}
\put(25,5){\vector(1,-1){2}}
\put(-120,30){\line(1,-1){30}}
\put(-90,30){\line(-1,-1){13}}
\put(-120,0){\line(1,1){13}}

\put(-70,10){\text{$\mathcal{R}$}}
\put(-86,10){\text{$=$}}

\put(-115,5){\vector(-1,-1){2}}
\put(-95,5){\vector(1,-1){2}}

\end{picture}
 \caption{Basic types of crossings}\label{crossings}
\end{figure}
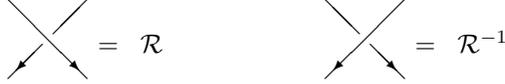

The choice of operators we associate with different types of crossings and turning points is made so that they obey ambient isotopy invariance. We also have to ensure that the operators satisfy the Reidemeister moves, which provides additional constrains.

The second Reidemeister move is satisfied because we use inverse operator $\mathcal{R}^{-1}$ in the second type of crossing. In order to satisfy the first and the third moves, the operators must obey the following equations:
\begin{align}
    \Tr_2 (\mathcal{R}^{\pm 1}(I\otimes \mathcal{W}))  & = I \label{1st}, \\
    (\r \otimes I)(I \otimes \r)(\r \otimes I) &  = (I\otimes \r)(\r \otimes I)(I \otimes \r) \label{YB1},
\end{align}
where $\mathcal{W}=\mathcal{M}^2$, $I$ is the identity operator. Solution to these equations is a pair of operators $(\mathcal{R},\mathcal{W})$ acting on the tensor product of representations (in our case, $\mathcal{U}_q(sl_N)$).  Eq.\ref{YB1} is called Yang-Baxter equation.

Strictly speaking, eq.\ref{1st} depends on the framing of the knot (we will discuss it a bit more in sec.\ref{frame}) and can be written as $ \Tr_2 (\mathcal{R}^{\pm 1}(I\otimes \mathcal{W})) =q^{\Omega_2}\, I$. It means that the answer can depend on the representation of the knot on the knot diagram and can differ by a scalar factor. In this paper, we normalise $\mathcal{W}$ so that it satisfies the eq.\ref{1st}.

Invariants of knots and links are given as contractions of operators corresponding to the elements of knot or link diagrams.

\subsubsection*{Links}
When one calculates invariants of links, one has an additional freedom: one can choose different representations for different components of the link. In our case, this means that we choose representations with different parameters $\l_i^{(1)}$, $\l_i^{(2)}, \, \dots$. In this case, one uses the \R-matrices that depend on two different colors $\r(\l^{(1)},\l^{(2)})$, and the Reidemeister moves give us the following equations
\begin{align}
    \Tr_2 (\mathcal{R}^{\pm 1}(\l,\l)(I\otimes \mathcal{W}))   & = I \label{1stMIX}, \\
   \r(\l^{(1)},\l^{(2)})\r^{-1}(\l^{(2)},\l^{(1)}) &  = I, \label{2ndMIX}\\
    (\r(\l^{(1)},\l^{(2)}) \otimes I)(I \otimes \r(\l^{(1)},\l^{(3)}))(\r(\l^{(2)},\l^{(3)}) \otimes I) &  =  \label{YBMIX}\\ (I\otimes \r(\l^{(2)},\l^{(3)}))(\r(\l^{(1)},\l^{(3)}) \otimes I)(I \otimes \r(\l^{(1)},\,&\l^{(2)}))\nn.
\end{align}

Eq.\ref{1stMIX} is one-colored because the corresponding Reidemeister move operates with one strand, eq.\ref{2ndMIX} defines the inverse \R-matrix,  eq.\ref{YBMIX} is the enhanced Yang-Baxter equation. These equations allow one to define the colored \R-matrix and the weight matrix $\W$, which allow one to calculate invariants of links.

\subsection{Braid representations}

\begin{figure}[H]
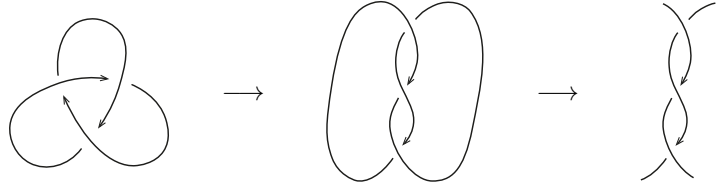

\bc
\trefoilzero \qquad $\longrightarrow$ \qquad \trefoilB \qquad $\longrightarrow$ \qquad \trefoiljust{}
\ec
  \caption{Representation of trefoil in the form of two-strand braid}
    \label{braidtrefoil}
\end{figure}

In this work, we use the RT approach for representations of knots and links as closed braids, which allows one to avoid turning points on the knot diagrams except for the closures of braids. The trefoil knot can be realized as a braid, for instance, as at the last picture of Fig.\ref{braidtrefoil}.

Each strand of the braid is associated with representation of $\mathcal{U}_q(sl_N)$ $T_i$, and one should use operators acting on the tensor product  $\bigotimes T_i$ of representations on each strand. The \R-matrix operator acts on two strands, and for three and more strand braids one uses
\begin{equation}
    \mathcal{R}_i = \overbrace{I \otimes I \otimes \dots}^{i-1} \otimes \mathcal{R} \otimes \overbrace{\dots \otimes I }^{s-i-1},
\end{equation}
where $I$ is the identity operator, $s$ is the number of strands in the braid. For the three-strand braid, one needs two $\mathcal{R}$-matrices: $\mathcal{R}_1 = \mathcal{R}\otimes I$, $ \mathcal{R}_2 = I\otimes\mathcal{R}$. To close the braid, one uses the quantum trace operation
\begin{equation}
    {\rm Tr}_q \,\, A  = {\rm Tr} \,\,A\,\, \overbrace{  \mathcal{W}\otimes \mathcal{W} \otimes \dots \otimes \mathcal{W}}^{s},
    \label{trdefold}
\end{equation}
 which is defined via the operator $\mathcal{W} = \mathcal{M}^2$.

\begin{figure}[H]
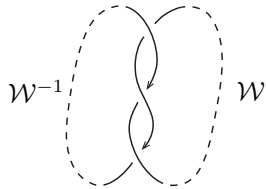

\begin{center}
      {\labellist
\pinlabel{$\mathcal{W}^{-1}$} at -20 65
\pinlabel{$\mathcal{W}$} at 127 65
\endlabellist
\trefoilBW
} \end{center}
\caption{Closure of a braid}
\end{figure}
One can get a polynomial invariant as the quantum trace of the product of the \R-matrices corresponding to crossings on the braid:
\begin{equation}
    \H^{\L} = \Tr_q \prod_i \r_i
\end{equation}

\begin{figure}[H]
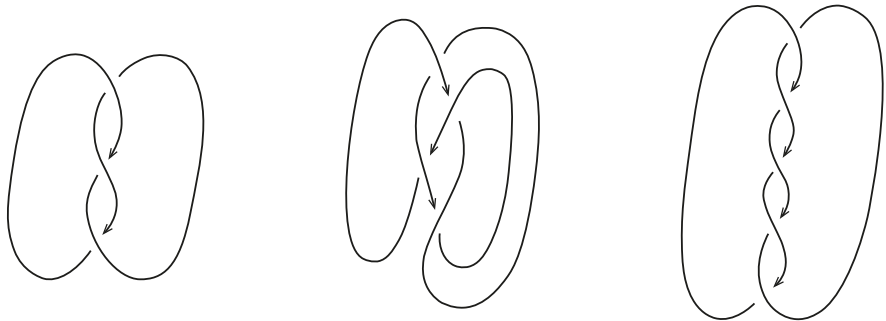

\bc
{\labellist
\endlabellist
\trefoil}
\quad \quad \quad \quad \quad
{\labellist
\endlabellist
\eight}
\quad \quad \quad \quad \quad
{\labellist
\endlabellist
\five}
\ec
\caption{Knots: trefoil (knot $3_1$), figure-eight knot (knot $4_1$) and knot $5_1$ in the braid form}
\label{knots}
\end{figure}

\begin{figure}[H]
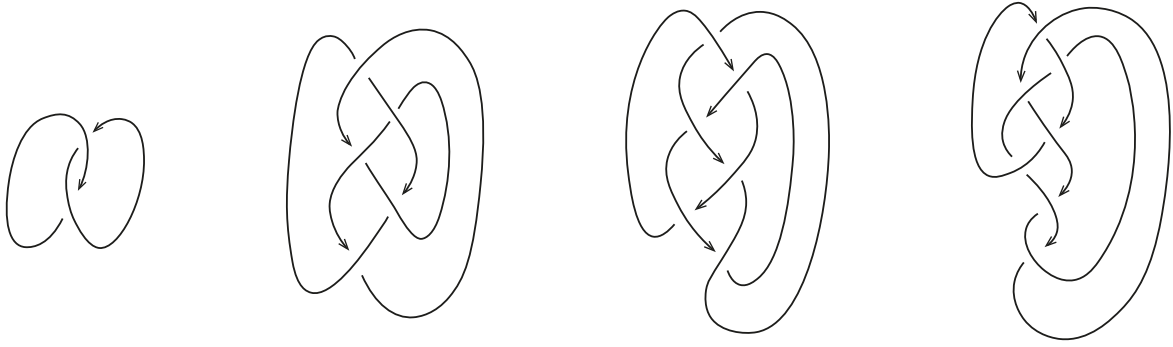

\bc
{\labellist
\endlabellist
\hopf}
\quad \quad \quad \quad \quad
{\labellist
\endlabellist
\whlink}
\quad \quad \quad \quad \quad
{\labellist
\endlabellist
\brlink}
\quad \quad \quad \quad \quad
{\labellist
\endlabellist
\lseven}
\ec
\caption{Links: the Hopf link (link $L_2 a_1$), the Whitehead link (link $L_5 a_1$), the Borromean rings (link $L_6 a_4$) and link $L_7 a_1$ in the braid form}
\label{links}
\end{figure}
To calculate the non-reduced (non-normalized) HOMFLY-PT polynomials $\H^{\L}$ of knots and links presented in Fig.\ref{knots} and Fig.\ref{links}, one can use the following formulas:
\begin{align}
      \H^{3_1} & = \Tr (\W^{-1}\otimes \W)\,\r^3, \nn \\
         \H^{4_1} & = \Tr (\W^{-1}\otimes \W\otimes \W)\,\r_1 \r_2^{-1} \r_1 \r_2^{-1}, \nn \\
         \H^{5_1} & = \Tr (\W^{-1}\otimes \W)\, \r^5. \nn \\
         & \nn \\
      \H^{L_2 a_1} & = \Tr (\W^{-1}(\l^{(1)})\otimes \W(\l^{(2)})) \, \r(\l^{(1)},\l^{(2)}) \r(\l^{(2)},\l^{(1)}) , \\
      \H^{L_5 a_1} & = \Tr ( \W^{-1}(\l^{(1)})\otimes \W(\l^{(2)}) \otimes  \W(\l^{(2)})) \nn \\ &\,\,\,\,\,\,\,\,\,\,\,\, \r_1^{-1}(\l^{(1)},\l^{(2)}) \r_2(\l^{(1)},\l^{(2)}) \r_1^{-1}(\l^{(2)},\l^{(2)}) \r_2(\l^{(2)},\l^{(1)}) \r_1^{-1}(\l^{(2)},\l^{(1)}) ,\nn \\
      \H^{L_6 a_4} & = \Tr ( \W^{-1}(\l^{(1)})\otimes \W(\l^{(2)}) \otimes  \W(\l^{(3)})) \nn  \\ &\,\,\,\,\,\,\,\,\,\,\,\, \r_1(\l^{(1)},\l^{(2)}) \r_2^{-1}(\l^{(1)},\l^{(3)}) \r_1(\l^{(2)},\l^{(3)}) \r_2^{-1}(\l^{(2)},\l^{(1)}) \r_1(\l^{(3)},\l^{(1)}) \r_2^{-1}(\l^{(3)},\l^{(2)}),\nn \\
      \H^{L_7 a_1} & = \Tr ( \W^{-1}(\l^{(1)})\otimes \W(\l^{(2)}) \otimes  \W(\l^{(2)}))  \nn  \\ &\,\,\,\,\,\,\,\,\,\,\,\, \r_1^{-1}(\l^{(1)},\l^{(2)}) \r_2(\l^{(1)},\l^{(2)}) \r_1^{-1}(\l^{(2)},\l^{(2)}) \r_2(\l^{(2)},\l^{(1)}) \r_1^{-1}(\l^{(2)},\l^{(1)}) \r_2^{2}(\l^{(2)},\l^{(2)}). \nn
\end{align}

\subsubsection{Framing of the knot and link polynomials \label{frame}}

Eqs.\ref{YB1}, \ref{2ndMIX} and \ref{YBMIX} as it is easy to see, do not fix in any way a general coefficient in front of $\mathcal{R}$-matrix, its normalization. In fact, there are several different choices of these normalizations. From the perspective of Chern-Simons theory, this corresponds to the fact that the observables which are studied actually correspond to a framed knot/link which looks more like a band and can have different number of twists \cite{MVA}.

There are several different natural choices of framing, depending on the situation, see \cite{Cab,BJM} for more details. We describe them as they look like for individual $\mathcal{R}$-matrices. The first choice called topological framing is natural for knots. It is given by eq.\ref{1st}, which can connect different diagrams of the same knot. This defines the normalization coefficient in front of the $\mathcal{R}$-matrix. This framing, however, cannot be defined when the $\mathcal{R}$-matrix corresponds to an intersection between components of a link with different parameters, as in eqs.\ref{2ndMIX} and \ref{YBMIX}, because for such intersections there is no eq.\ref{1st}.

Another choice comes from quantum groups. The product of irreducible representations of quantum groups can be expanded into a sum of other irreducible representations. This in part leads to a cabling procedure \cite{Cab}, which connects knot polynomials in different representations with each other. Using this procedure, one can rewrite the $\mathcal{R}$-matrix of the higher representation as a product of $\mathcal{R}$-matrices of lower representations. However this puts a certain restriction on the normalization of the $\mathcal{R}$-matrices, which are related by the cabling procedure. This framing is called vertical framing.

Finally, the third natural choice comes from Chern-Simons theory\footnote{In fact, there are more natural framings, for instance, the differential framing \cite{BJM}, however, we do not discuss them here.}. It is defined by the knot or link having zero linking number for a framed knot band. It is called canonical framing \cite{MVA}. Its defining characteristic is given via expansion of the knot or link polynomial: if one expands the HOMFLY-PT polynomial in $\hbar$, substituting $q=exp(i\hbar)$, then the linear in $\hbar$ term of the expansion should vanish. For knots, this framing actually coincides with the topological one. However, for links it is different.

In this paper, we study $\mathcal{R}$-matrices in the topological framing, since it can be most easily defined for the $\U_q(sl_N)$. Thus answers for the HOMFLY-PT polynomials should also be considered in the topological framing.

\subsection{Application of RT method to representations at roots of unity}
In order to use the RT approach, one needs to define the \R-matrices and the weight matrices for special representations that arise in the root-of-unity case. And naturally we want to use the same operators as we did in the not-a-root-of-unity case.

The problem arises because unreduced HOMFLY-PT polynomials that we get with the RT method from representations of $\mathcal{U}_q(sl_N)$ are proportional to the polynomial of unknot --- quantum dimension of representation, which is equal to Schur polynomials of representations in a special point. And it is always zero when we take into account  that $q$ is the $2m$-th root of unity. This problem is solved with calculation of reduced polynomials, which can be conducted with the modified quantum trace operation, which is equivalent to calculation of invariants of $(1,1)$-tangles.

\subsubsection*{Quantum trace}
\begin{figure}[H]
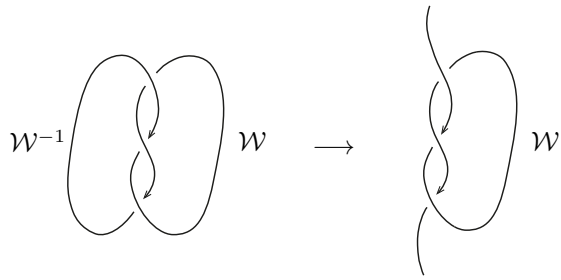

\bc
{\labellist
\pinlabel{$\mathcal{W}^{-1}$} at -20 65
\pinlabel{$\mathcal{W}$} at 127 65
\endlabellist
\trefoilB}
\qquad
\,\,\,\,
{\raisebox{0\height}{$\longrightarrow$}}
\qquad
{\labellist
\pinlabel{$\mathcal{W}$} at 88 92
\endlabellist
\trefoilBWcut}
\ec
\caption{Diagrams of the trefoil and the corresponding (1,1)-tangle}
\label{cutproc}
\end{figure}

The existence of one-to-one correspondence between knots/links and $(1,1)$-tangles \cite{tangles} allows one to calculate invariants of tangles and, in fact, to get reduced invariants of corresponding knots and links. It means that one should tear up one closure as in Fig.\ref{cutproc} and apply the same RT procedure which we described above.

In this case, we have one in-going and one out-going lines, this is why it is called $(1,1)$-tangle. The Reidemeister moves can be applied to such tangles if one fixes open lines. An invariant of the tangle is a diagonal constant matrix proportional to the reduced HOMFLY-PT polynomial of the corresponding knot or link $H^{(1,1)\text{-tangle}} = I\,\, H^{\L}$.

Equivalently, one can redefine the quantum trace operation and apply it to knots and links, not to tangles. In the definition, we omit one weight matrix $\W$, which effectively means that we cut one line in the diagram:
    \begin{equation}
  \boxed{  {\rm \Tr}^*_q \,\, A = {\rm Tr}\,\, A \,\,I \otimes \overbrace{ \W \otimes \dots \otimes \W}^{s-1}, }
    \label{trdef}
\end{equation}
where $I$ is the identity operator. In this case, we get an additional normalization coefficient, which is equal to classical dimension of the representation.

\begin{figure}[H]
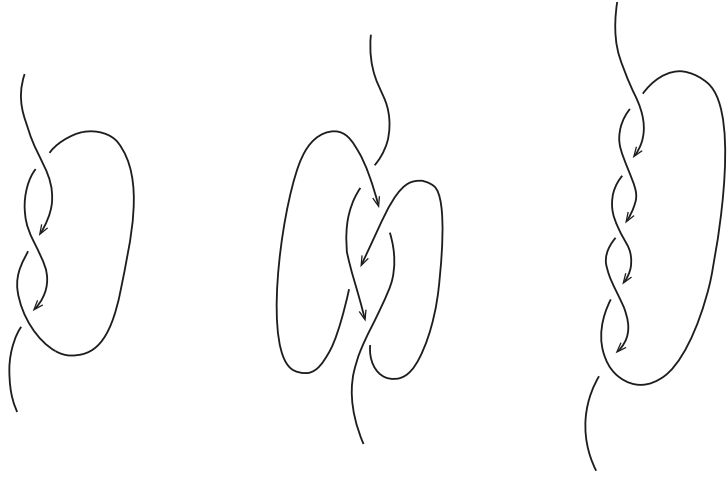

\bc
{\labellist
\endlabellist
\trefoilcut}
\quad \quad \quad \quad \quad
{\labellist
\endlabellist
\eightcut}
\quad \quad \quad \quad \quad
{\labellist
\endlabellist
\fivecut}
\ec
\caption{$(1,1)$-tangles corresponding to the trefoil, the figure-eight knot and knot $5_1$}
\label{knotscut}
\end{figure}

\begin{figure}[H]
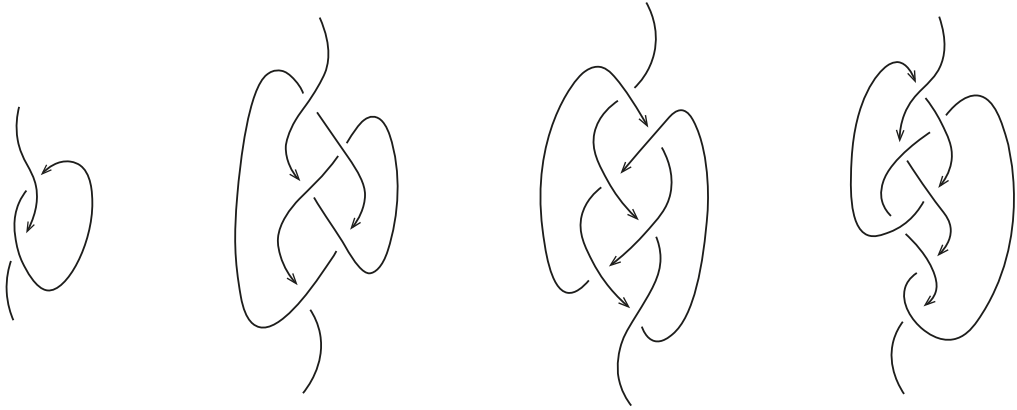

\bc
{\labellist
\endlabellist
\hopfcut}
\quad \quad \quad \quad \quad
{\labellist
\endlabellist
\whlinkcut}
\quad \quad \quad \quad \quad
{\labellist
\endlabellist
\brlinkcut}
\quad \quad \quad \quad \quad
{\labellist
\endlabellist
\lsevencut}
\ec
\caption{$(1,1)$-tangles corresponding to the Hopf link, the Whitehead link, the Borromean rings and link $L_7 a_1$}
\label{knotscut}
\end{figure}

One can use the following formulas to get invariants  at roots of unity:

\begin{align}
      P^{3_1} & = \Tr (I\otimes \W)\,\r^3, \nn \\
         P^{4_1} & = \Tr (\W^{-1}\otimes I \otimes \W)\,\r_1 \r_2^{-1} \r_1 \r_2^{-1}, \nn \\
         P^{5_1} & = \Tr (I \otimes \W)\, \r^5. \nn \\
         & \nn \\
      P^{L_2 a_1} & = \Tr (I \otimes \W(\l^{(2)})) \, \r(\l^{(1)},\l^{(2)}) \r(\l^{(2)},\l^{(1)}) ,\label{ppolynomial} \\
      P^{L_5 a_1} & = \Tr ( \W^{-1}(\l^{(1)})\otimes I \otimes  \W(\l^{(2)})) \nn \\ &\,\,\,\,\,\,\,\,\,\,\,\, \r_1^{-1}(\l^{(1)},\l^{(2)}) \r_2(\l^{(1)},\l^{(2)}) \r_1^{-1}(\l^{(2)},\l^{(2)}) \r_2(\l^{(2)},\l^{(1)}) \r_1^{-1}(\l^{(2)},\l^{(1)}) ,\nn \\
      P^{L_6 a_4} & = \Tr ( \W^{-1}(\l^{(1)})\otimes I \otimes  \W(\l^{(3)})) \nn  \\ &\,\,\,\,\,\,\,\,\,\,\,\, \r_1(\l^{(1)},\l^{(2)}) \r_2^{-1}(\l^{(1)},\l^{(3)}) \r_1(\l^{(2)},\l^{(3)}) \r_2^{-1}(\l^{(2)},\l^{(1)}) \r_1(\l^{(3)},\l^{(1)}) \r_2^{-1}(\l^{(3)},\l^{(2)}),\nn \\
      P^{L_7 a_1} & = \Tr ( \W^{-1}(\l^{(1)})\otimes I \otimes  \W(\l^{(2)}))  \nn  \\ &\,\,\,\,\,\,\,\,\,\,\,\, \r_1^{-1}(\l^{(1)},\l^{(2)}) \r_2(\l^{(1)},\l^{(2)}) \r_1^{-1}(\l^{(2)},\l^{(2)}) \r_2(\l^{(2)},\l^{(1)}) \r_1^{-1}(\l^{(2)},\l^{(1)}) \r_2^{2}(\l^{(2)},\l^{(2)}). \nn
\end{align}

\section{Invariants corresponding to nilpotent representations of $\U_q(sl_2)$\label{secSl2}}
\subsection{Nilpotent representations of $\U_q(sl_2)$ with parameter \label{repsSl2}}
In this section, we start with the simplest case of $\U_q(sl_2)$, and consider the nilpotent representation $W_{m,2}(q^{\mu})$ with a continuous parameter $q^{\mu} = \l$. $W_{m,2}(q^{\mu})$ is an irreducible representation on the vector space $\mathcal{V}_m$ of dimension $m$ with basis vectors $v_i$, $i={0,\dots, m-1}$. The generators are $E$, $F$ and $K$. The discrete parameters are $c_{jl} = 1$ and $\eta_{ijl}=1$. $p_{22}=0$, $p_{11} = p_{12} = \mu+1$, which gives one continuous parameter.
The highest weight vector $v_0$ corresponds to the state
\begin{equation}
   v_0 =
  \left |  \begin{array}{ccc}
         \mu+1 & & 0 \\
         & \mu+1 &\\
    \end{array} \right \rangle,
\end{equation}
and
\begin{equation}
   v_i =
  \left |  \begin{array}{ccc}
         \mu+1 & & 0 \\
         & \mu+1-i &\\
    \end{array} \right \rangle.
\end{equation}
Then the matrices of representations acting on the basis vectors $v_i$ are the following:

\begin{equation}
    \begin{array}{lll}
          K v_i= q^{\mu -2i} v_i, \qquad  & E v_i = [i][\mu-i+1] v_{i-1},\qquad & F v_i = v_{i+1}, \\
     & E v_0 = 0, & F v_{m-1} = 0. \\
     \label{gensl2}
    \end{array}
\end{equation}
 Here $K v_0 = q^{\mu} v_0$, which means that $\mu$ plays the role of the highest weight. It is convenient for the computations to make the substitution $q^\mu = \lambda$, polynomial invariants associated with these representations that we calculated depend on $\lambda$.

\subsection{\R-matrix and weight matrix \label{Rsl2}}
We use the explicit formula for the universal \R-matrix (\ref{UniR}) and choose the normalization coefficient to fix the topological framing:

\begin{equation}
    \r_{m,2} = \left.  q^{-\mu ^2/2}q^{\mu(m-1)} \r_u \right|_{sl_2},
\end{equation}

\begin{dmath}
   \mathcal{R}_{m,2} \, (v_i \otimes v_j) =  \sum_{n=0}^{m-1} q^{\mu(m-1-i-j)+2(i-n)(j+n)+\frac{n(n-1)}{2}} \frac{(q-q^{-1})^n}{[n]!} \\  {[i-n+1;n]} {[\mu-i+n;n]} \, {(v_{j+n}\otimes v_{i-n})},
   \label{R2}
\end{dmath}

where $[x;n] = \prod_{i=0}^{n-1}[x-i]$. The weight matrix $\W_{m,2}$ is defined via the first Reidemeister move (\ref{1st}) and is equal to

\begin{equation}
    \mathcal{W}_{m,2} \, v_i =q^{-\mu (m-1)-2i} v_i =  q^{-\mu\, m} K v_i.
    \label{W2}
\end{equation}

To calculate invariants of links, one needs to write these formulas with two colors:
\begin{equation}
    \r_{m,2}^{(1),(2)}=  \left. q^{-\frac{1}{2}\mu^{(1)} \mu^{(2)} } \left(q^{\mu^{(1)}}q^{\mu^{(2)}} \right)^{(m-1)/2} \r_u \right|_{sl_2},
\end{equation}

\begin{dmath}
   \mathcal{R}_{m,2}^{(1),(2)} \, (v_i^{\mu^{(1)}} \otimes v_j^{\mu^{(2)}}) =   \sum_{n=0}^{m-1} q^{\m^{(2)}(m-1-i+n)-\mu^{(1)}(j+n)+2(i-n)(j+n)+\frac{n(n-1)}{2}} \frac{(q-q^{-1})^n}{[n]!} \\  {[i-n+1;n]} {[\mu^{(1)}-i+n;n]} \, (v_{j+n}^{\m^{(2)}}\otimes v_{i-n}^{\m^{(1)}}).
   \label{R2mix}
\end{dmath}

\subsection{Invariants of knots and links}
The reduced polynomials of knots and links can be calculated by formula (\ref{ppolynomial}) with $\r_{m,2}$ from (\ref{R2}) and $\W_{m,2}$ from (\ref{W2}). In order to get invariants of knots and links, one should use normalization coefficient (\ref{norm}), which, in this case, is
\begin{equation}
    \Xi_{m,2}(\l) = \xi_m(\lambda) =\prod_{i=0}^{m-2} \{\lambda q^{-i}\}, \end{equation}
Then, the invariants $\P_{m,2}^{\L}(\l^{(k)}_i)$ (where $k$ numerates the components of link, $\l^{(1)}_i$ is color of the open component) of knots and links at roots of unity are
\begin{equation}
\P_{m,2}^{\L}(\l^{(k)}_i) = {P_{m,2}^{\L}(\l^{(k)}_i) \over \prod_{i=0}^{m-2} \{\lambda^{({1})}_i q^{-i}\}}.
\end{equation}

\subsection{Connection with HOMFLY-PT and Alexander polynomials}
The reduced polynomials $P_{m,N}^{\L}(\l^i)$ coincide with the reduced HOMFLY-PT polynomials in symmetric representations $[m-1]$, when parameters $\l^i$ equal to the highest weights of representations $[m-1]$:
\begin{equation}
    P_{m,2}^{\L}(q, \l^i = q^{m-1}) = H_{[m-1]}^{\L}(A=q^2, q)
\end{equation}
Another connection is that $P_{2,2}^{\K}$ coincide with the Alexander polynomials of knots

\begin{equation}
    P_{2,2}^{\K}(\l) = \A^{\K}(\l)
\end{equation}
while $\P_{2,2}^{\L}$, with the multivariable Alexander polynomials of links
\begin{equation}
    \P_{2,2}^{\L}(\l^{(1)},\l^{(2)},\dots) = \D^{\L}(\l^{(1)},\l^{(2)},\dots)
\end{equation}

\section{Knot and link invariants corresponding to nilpotent representations of $\U_q(sl_3)$ and $\U_q(sl_4)$ \label{secSl34}}
\subsection{Nilpotent representations of $\U_q(sl_3)$ with parameters \label{repsSl3}}
In this section, we consider the nilpotent representation $W_{m,3}(q^{\mu_1},q^{\mu_2})$ of \uqslthree{} with continuous parameters. $W_{m,3}(q^{\mu_1},q^{\mu_2})$  is an irreducible representation on the vector space $\mathcal{V}_{m^3}$ of dimension $m^{3}$. The generators are $E_1$, $E_2$, $F_1$, $F_2$, $K_1$, $K_2$. The discrete parameters are $c_{jl} = 1$, $\eta_{ijl}=1$. There are two continuous parameters $\mu_1$ and $\mu_2$, which enter the parameters of the highest weight state in the following way: $p_{33}=0$, $p_{23} = p_{22} =\mu_2+1$,  $p_{13} = p_{12}=p_{11} = \mu_1+\mu_2+2$. The highest vector corresponds to the state
\begin{equation}
v_0 \,\,\,= \,\,\,
\left |
    \begin{array}{ccccc}
    \mu_1+\mu_2+2 & & \mu_2+1 & & 0 \\
    & \mu_1+\mu_2+2 & & \mu_2+1 & \\
    && \mu_1+\mu_2+2 && \\
    \end{array}
    \right \rangle,
\end{equation}
and we use the following notation for other vectors of the vector space $\mathcal{V}_{m^3}$:
\begin{equation}
v(n_1,n_2,n_3) \,\,\,= \,\,\,
\left |
    \begin{array}{ccccc}
    \mu_1+\mu_2+2 & & \mu_2+1 & & 0 \\
    & \mu_1+\mu_2+2-n_2 & & \mu_2+1-n_3 & \\
    && \mu_1+\mu_2+2-n_1 && \\
    \end{array}
    \right \rangle
\end{equation}
The operators $E_1$, $E_2$, $F_1$, $F_2$ are
\begin{align}
      K_1 \, v(n_1,n_2,n_3) \,=\, & q^{\mu_1-2n_1+n_2+n_3} \, v(n_1,n_2,n_3), \\
      K_2 \, v(n_1,n_2,n_3) \,=\, & q^{\mu_2+n_1-2n_2-2n_3}\, v(n_1,n_2,n_3),\nn \\ \nn\\
       E_1 \,v(n_1,n_2,n_3) \,=\, & [n_1-n_2][\mu_1+1+n_3-n_1]\,v(n_1+1,n_2,n_3), \nn \\
        E_2 \,v(n_1,n_2,n_3) \,=\, & [n_2][\mu_1+1-n_2][\mu_1+\mu_2+1-n_2][\mu_1+2+n_3-n_2]^{-1/2}[\mu_1+1+n_3-n_2]^{-1/2} v(n_1,n_2+1,n_3)+ \nn \\
        \text{    }  & [\mu_1+1+n_3][n_3][\mu_2+1-n_3][\mu_1+n_3-n_2]^{-1/2}[\mu_1+1+n_3-n_2]^{-1/2}\, v(n_1,n_2,n_3+1) ,\nn\\
         F_1 \,v(n_1,n_2,n_3) \,=\, & v(n_1-1, n_2,n_3), \nn \\
         F_2\, v(n_1,n_2,n_3)\, =\, & [n_2-n_1] [\mu_1+1+n_3-n_2]^{-1/2} [\mu_1+n_3-n_2]^{-1/2}\,v(n_1,n_2-1,n_3) + \nn \\
        \text{    }  & [\mu_1+1+n_3-n_1] [\mu_1+1+n_3-n_2]^{-1/2} [\mu_1+2+n_3-n_2]^{-1/2} \,v(n_1,n_2,n_3-1). \nn
\end{align}
The choice of continuous parameters $\mu_1$ and $\mu_2$ in representations is made so that these parameters play the roles of the highest weights: $K_1 v_0 = q^{\mu_1} v_0$, $K_2 v_0 = q^{\mu_2} v_0$. It is also convenient to use the substitutions: $q^{\mu_1} = \l_1$ and $q^{\mu_2} = \l_2$, which enter invariants associated with representations of $\U_q(sl_3)$.

\bigskip

In this paper, due to computational restrictions, we considered three representations of $\U_q(sl_3)$ that correspond to the $4$-th, $6$-th and $8$-th roots of unity.
When $m=2$, representation $W_{2,3}(q^{\mu_1}, q^{\mu_2})$ acts on the 8-dimensional vector space that consists of the following vectors:
\begin{equation}
    \begin{array}{llll}
        v_0 = v(0,0,0), \,\,\,\,\,\,\,\,\,\, & v_2 = v(0,0,1), \,\,\,\,\,\,\,\,\,\, & v_4=v(1,0,1) , \,\,\,\,\,\,\,\,\,\, & v_6=v(1,1,0), \\
          v_1 = v(1,0,0), \,\,\,\,\,\,\,\,\,\, & v_3 = v(1,1,0), \,\,\,\,\,\,\,\,\,\, & v_5=v(2,1,0) , \,\,\,\,\,\,\,\,\,\, & v_7=v(2,1,1). \\
    \end{array}
\end{equation}
When  $m=3$, we get the representation $W_{3,3}(q^{\mu_1},q^{\mu_2})$ acting on the 27-dimensional vector space with the vectors:
\begin{equation}
    \begin{array}{llll}
        v_0 = v(0,0,0), \,\,\,\,\,\,\,\,\,\, & v_7 = v(2,1,0), \,\,\,\,\,\,\,\,\,\, & v_{14}=v(2,1,1) , \,\,\,\,\,\,\,\,\,\, & v_{21}=v(3,2,1), \\
          v_1 = v(1,0,0), \,\,\,\,\,\,\,\,\,\, & v_8 = v(2,0,1), \,\,\,\,\,\,\,\,\,\, & v_{15}=v(1,1,2) , \,\,\,\,\,\,\,\,\,\, & v_{22}=v(3,1,2), \\
          v_2 = v(0,0,1), \,\,\,\,\,\,\,\,\,\, & v_9 = v(1,1,1), \,\,\,\,\,\,\,\,\,\, & v_{16}=v(3,2,0) , \,\,\,\,\,\,\,\,\,\, & v_{23}=v(2,2,2), \\
          v_3 = v(2,0,0), \,\,\,\,\,\,\,\,\,\, & v_{10} = v(1,0,2), \,\,\,\,\,\,\,\,\,\, & v_{17}=v(3,1,1) , \,\,\,\,\,\,\,\,\,\, & v_{24}=v(4,2,1), \\
          v_4 = v(1,1,0), \,\,\,\,\,\,\,\,\,\, & v_{11} = v(3,1,0), \,\,\,\,\,\,\,\,\,\, & v_{18}=v(2,2,1) , \,\,\,\,\,\,\,\,\,\, & v_{25}=v(3,2,2), \\
          v_5 = v(1,0,1), \,\,\,\,\,\,\,\,\,\, & v_{12} = v(2,0,2), \,\,\,\,\,\,\,\,\,\, & v_{19}=v(2,1,2) , \,\,\,\,\,\,\,\,\,\, & v_{26}=v(4,2,2), \\
          v_6 = v(1,0,2), \,\,\,\,\,\,\,\,\,\, & v_{13} = v(2,2,0), \,\,\,\,\,\,\,\,\,\, & v_{20}=v(4,2,0) . \,\,\,\,\,\,\,\,\,\, &  \\
    \end{array}
\end{equation}
In order to avoid irrationalities (denominators containing roots) in our computations, we made a rational rescaling of operators of the representation  with the diagonal matrix $M_6$: $\Tilde{E}_{1,2} = M_6 E_{1,2} M_6^{-1}$, $\Tilde{F}_{1,2} = M_6 F_{1,2} M_6^{-1}$, $\Tilde{K}_{1,2} = K_{1,2}$, where

\begin{equation}
    \begin{array}{ll}
        M_6 \,\,=\,\, & {\rm diag} \left(
\sqrt{[\mu_1 +1]},\sqrt{[\mu_1 +1]},\sqrt{[\mu_1 +2]},\sqrt{[\mu_1 +1]},\sqrt{[\mu_1 ]} \right.,\sqrt{[\mu_1
   +2]},\sqrt{[\mu_1 +3]}, \sqrt{[\mu_1 ]},\sqrt{[\mu_1 +2]},\\& \sqrt{[\mu_1 +1]},\sqrt{[\mu_1 +3]},\sqrt{[\mu_1
   ]},\sqrt{[\mu_1 -1]},\sqrt{[\mu_1 +3]},\sqrt{[\mu_1 +1]}, \sqrt{[\mu_1 +2]},\sqrt{[\mu_1 -1]},\sqrt{[\mu_1
   +1]},\sqrt{[\mu_1 ]} \\ & \sqrt{[\mu_1 +2]},\sqrt{[\mu_1 -1]},\sqrt{[\mu_1 ]},\sqrt{[\mu_1 +2]},\sqrt{[\mu_1
   +1]},\sqrt{[\mu_1 ]}\left.,\sqrt{[\mu_1 +1]},\sqrt{[\mu_1 +1]} \right) \\
    \end{array}
\end{equation}

When  $m=4$, we get representation $W_{4,3}(q^{\mu_1},q^{\mu_2})$ acting on the 64-dimensional vector space with the vectors:

\begin{equation}
   \begin{array}{llll}
        v_{0} = v(0, 0, 0),\,\,\,\,\,\,\,\,\,\, & v_{1} = v(1, 0, 0),\,\,\,\,\,\,\,\,\,\,& v_{2} = v(0, 0, 1),\,\,\,\,\,\,\,\,\,\,& v_{3} = v(2, 0, 0), \\
        v_{4} = v(1, 0,  1), & v_{5} = v(1, 1, 0), &  v_{6} = v(0, 0, 2), & v_{7} = v(3, 0, 0), \\
        v_{8} = v(2, 0, 1), & v_{9} = v(2,  1, 0), & v_{10} = v(1, 0, 2), &  v_{11} = v(1, 1, 1), \\
        v_{12} = v(0, 0, 3), & v_{13} = v(3, 0, 1), &  v_{14} = v(3, 1, 0), & v_{15} = v(2, 0, 2), \\
        v_{16} = v(2, 1, 1), & v_{17} = v(1, 0, 3), &  v_{18} = v(2, 2, 0),& v_{19} = v(1, 1, 2), \\
        v_{20} = v(4, 1, 0), & v_{21} = v(3, 0, 2), & v_{22} = v(3, 1, 1), & v_{23} = v(2, 0, 3), \\
        v_{24} = v(3, 2, 0), & v_{25} = v(2, 1, 2), & v_{26} = v(2, 2, 1),& v_{27} = v(1, 1, 3), \\
        v_{28} = v(4, 1, 1), & v_{29} = v(3, 0, 3), & v_{30} = v(4, 2, 0), & v_{31} = v(3, 1, 2), \\
        v_{32} = v(3, 2, 1), & v_{33} = v(2, 1, 3),&  v_{34} = v(3, 3, 0),& v_{35} = v(2, 2, 2),\\
        v_{36} = v(5,  2, 0), & v_{37} = v(4, 1, 2),& v_{38} = v(4, 2, 1),& v_{39} = v(3, 1, 3),\\
        v_{40} = v(4, 3, 0),&  v_{41} = v(3, 2, 2), & v_{42} = v(3, 3, 1), & v_{43} = v(2, 2, 3),\\
        v_{44} = v(5, 2, 1),&  v_{45} = v(4, 1, 3), & v_{46} = v(5, 3, 0),&  v_{47} = v(4, 2, 2),\\
        v_{48} = v(4, 3, 1), & v_{49} = v(3, 2, 3), & v_{50} = v(3, 3, 2), & v_{51} = v(6, 3, 0), \\
        v_{52} = v(5, 2, 2),& v_{53} = v(5, 3, 1),& v_{54} = v(4, 2, 3),& v_{55} = v(4, 3, 2),\\
        v_{56} = v(3, 3, 3), &v_{57} = v(6, 3, 1),& v_{58} = v(5, 2,3),&  v_{59} = v(5, 3, 2),\\
        v_{60} = v(4, 3, 3),& v_{61} = v(6, 3, 2),& v_{62} = v(5, 3, 3),& v_{63} = v(6, 3, 3) \\
   \end{array}
\end{equation}

In order to avoid irrationalities (denominators containing roots) in our computations, we made a rational rescaling of operators of the representation  with the diagonal matrix $M_8$: $\Tilde{E}_{1,2} = M_8 E_{1,2} M_8^{-1}$, $\Tilde{F}_{1,2} = M_8 F_{1,2} M_8^{-1}$, $\Tilde{K}_{1,2} = K_{1,2}$, where

\begin{equation}
    \begin{array}{ll}
        M_8 \,\,=\,\, & {\rm diag} \left( \sqrt{  [\mu_1 +1]},\sqrt{  [\mu_1 +1]},\right. \sqrt{  [\mu_1 +2]},\sqrt{  [\mu_1 +1]},\sqrt{  [\mu_1
   +2]},\sqrt{  [\mu_1 ]},\sqrt{  [\mu_1 +3]},\sqrt{  [\mu_1 +1]},\sqrt{  [\mu_1 +2]},\\&  \sqrt{  [\mu_1 ]},\sqrt{  [\mu_1
   +3]},\sqrt{  [\mu_1 +1]},\sqrt{  [\mu_1 +4]},\sqrt{  [\mu_1 +2]},\sqrt{  [\mu_1 ]},\sqrt{  [\mu_1
   +3]},\sqrt{  [\mu_1 +1]},\sqrt{  [\mu_1 +4]},\sqrt{  [\mu_1 -1]},\\&  \sqrt{  [\mu_1 +2]},\sqrt{  [\mu_1
   ]},\sqrt{  [\mu_1 +3]},\sqrt{  [\mu_1 +1]},\sqrt{  [\mu_1 +4]},\sqrt{  [\mu_1 -1]},\sqrt{  [\mu_1
   +2]},\sqrt{  [\mu_1 ]},\sqrt{  [\mu_1 +3]},\sqrt{  [\mu_1 +1]},\\& \sqrt{  [\mu_1 +4]},\sqrt{  [\mu_1
   -1]},\sqrt{  [\mu_1 +2]},\sqrt{  [\mu_1 ]},\sqrt{  [\mu_1 +3]},\sqrt{  [\mu_1 -2]},\sqrt{  [\mu_1
   +1]},\sqrt{  [\mu_1 -1]},\sqrt{  [\mu_1 +2]},\sqrt{  [\mu_1 ]},\\& \sqrt{  [\mu_1 +3]},\sqrt{  [\mu_1
   -2]},\sqrt{  [\mu_1 +1]},\sqrt{  [\mu_1 -1]},\sqrt{  [\mu_1 +2]},\sqrt{  [\mu_1 ]},\sqrt{  [\mu_1
   +3]},\sqrt{  [\mu_1 -2]},\sqrt{  [\mu_1 +1]},\\& \sqrt{  [\mu_1 -1]},\sqrt{  [\mu_1 +2]},\sqrt{  [\mu_1
   ]},\sqrt{  [\mu_1 -2]},\sqrt{  [\mu_1 +1]},\sqrt{  [\mu_1 -1]},\sqrt{  [\mu_1 +2]},\sqrt{  [\mu_1 ]},\sqrt{  [\mu_1
   +1]},\sqrt{  [\mu_1 -1]},\\& \sqrt{  [\mu_1 +2]},\sqrt{  [\mu_1 ]},\sqrt{  [\mu_1 +1]},\sqrt{  [\mu_1 ]},\sqrt{  [\mu_1
   +1]},\left. \sqrt{  [\mu_1 +1]}  \right ) \\
    \end{array}
\end{equation}

\subsection{Nilpotent representations of $\U_q(sl_4)$ with parameters \label{repsSl4}}
In this section, we consider representation $W_{2,4}(q^{\mu_1},q^{\mu_2},q^{\mu_3})$ of \uqslfour{} with continuous parameters. The dimension of these representations is $m^6$. The generators are $E_1$, $E_2$, $E_3$, $F_1$,  $F_2$, $F_3$, $K_1$, $K_2$, $K_3$. There are three continuous parameters: $\mu_1$, $\mu_2$, $\mu_3$ or $\l_1 = q^{\mu_1}$, $\l_2 = q^{\mu_2}$, $\l_3 = q^{\mu_3}$. $p_{1,i}= \mu_1+\mu_2+\mu_3+3$, $p_{2,i} =  \mu_2+ \mu_3+2$, $p_{3,i} =  \mu_3 +1$ and $p_{4,4} = 0$. The highest weight vector is

\begin{equation}
v_0 \,\,\,= \,\,\,
\left |
    \begin{array}{ccccccc}
  \mu_1+\mu_2+\mu_3+3 &&  \mu_2+ \mu_3+2 & & \mu_3 +1 & & 0 \\
 &   \mu_1+\mu_2+\mu_3+3 &&  \mu_2+ \mu_3+2 & & \mu_3 +1 &  \\
 &&  \mu_1+\mu_2+\mu_3+3 &&  \mu_2+ \mu_3+2 & &  \\
&&&   \mu_1+\mu_2+\mu_3+3 &&& \\
    \end{array}
    \right. \miniket{} \, ,
\end{equation}
and we use the following notation for other vectors of the vector space $\mathcal{V}_{m^6}$:
\begin{eqnarray}
v(n_1,n_2,n_3,n_4,n_5,n_6) \,\,\,= \,\,\,\,\,\,\,\,\,\,\,\,\,\,\,\,\,\,\,\,\,\,\,\,\,\,\,\,\,\,\,\,\,\,\,\,\,\,\,\,\,\,\,\,\,\,\,\,\,\,\,\,\,\,\,\,\,\,\,\,\,\,\,\,\,\,\,\,\,\,\,\,\,\,\,\,\,\,\,\,\,\,\,\,\,\,\,\,\,\,\,\,\,\,\,\,\,\,\,\,\,\,\,\,\,\,\,\,\,\,\,\,\,\,\,\,\,\,\,\,\,\,\,\,\,\, \,\,\,\,\,\,\,\,\,\,\,\,\,\,\,\,\,\,\,\,\,\,\,\,\,\,\,\,\,\,\,\,\,\,\,\,\,\,\,\,\,\,\,\,\,\,\,\,\,\,\,\,\,\,\,\,\,\,\,\,\,\,\,\,\,\,\,\,\,\,\,\,\,\,\,\,\,\,\,\,\,\,\,\,\,\,\,\,\,\,\,\,\,\,\,\,\,\,\,\,\,\,\,\,\,\,\,\\
{\small
\left |
    \begin{array}{ccccccc}
  \mu_1+\mu_2+\mu_3+3 &&  \mu_2+ \mu_3+2 & & \mu_3 +1 & & 0 \\
 &   \mu_1+\mu_2+\mu_3+3-n_4 &&  \mu_2+ \mu_3+2-n_5 & & \mu_3 +1-n_6 &  \\
 &&  \mu_1+\mu_2+\mu_3+3-n_2 &&  \mu_2+ \mu_3+2-n_3 & &  \\
&&&   \mu_1+\mu_2+\mu_3+3-n_1 &&& \nn \\
    \end{array}
   \right. \miniket{} \, }
\end{eqnarray}
Representation $W_{2,4}(q^{\mu_1},q^{\mu_2},q^{\mu_3})$ is the 64-dimensional representation with the following basis vectors
\begin{equation}
    \begin{array}{llll}
 v_{0} = v(0, 0, 0, 0, 0, 0),\,\,\,\,\,\,\,\,\,\, &  v_{1} = v(1, 0, 0, 0, 0, 0),\,\,\,\,\,\,\,\,\,\,& v_{2} = v(0, 0, 1, 0, 0, 0),\,\,\,\,\,\,\,\,\,\,& v_{3} = v(0, 0, 0, 0, 0, 1),\\
 v_{4} = v(1, 0, 1, 0, 0, 0), & v_{5} = v(1, 0, 0, 0, 0,1),& v_{6} = v(1, 1, 0, 0, 0, 0),& v_{7} = v(0, 0, 1, 0, 0, 1),\\
 v_{8} = v(0, 0, 1, 0, 1, 0),& v_{9} = v(2, 1, 0, 0, 0, 0),& v_{10} = v(1, 0, 1, 0, 0, 1),& v_{11} = v(1, 0, 1, 0,1, 0),\\
 v_{12} = v(1, 1, 1, 0, 0, 0),& v_{13} = v(1, 1, 0, 0, 0, 1),& v_{14} = v(0, 0, 2, 0, 1, 0),& v_{15} = v(1, 1, 0, 1, 0, 0),\\
 v_{16} = v(0, 0, 1, 0, 1, 1),& v_{17} = v(2, 1, 1,0, 0, 0),& v_{18} = v(2, 1, 0, 0, 0, 1),& v_{19} = v(1, 0, 2, 0, 1, 0),\\
 v_{20} = v(2, 1, 0, 1, 0, 0),& v_{21} = v(1, 0, 1, 0, 1, 1),& v_{22} = v(1, 1, 1, 0, 0, 1),& v_{23} = v(1, 1,1, 0, 1, 0), \\
 v_{24} = v(1, 1, 1, 1, 0, 0),& v_{25} = v(0, 0, 2, 0, 1, 1),& v_{26} = v(1, 1, 0, 1, 0, 1),& v_{27} = v(2, 1, 1, 0, 0, 1),\\
 v_{28} = v(2, 1, 1, 0, 1, 0),& v_{28} = v(2,1, 1, 1, 0, 0),& v_{30} = v(1, 0, 2, 0, 1, 1),& v_{31} = v(2, 1, 0, 1, 0, 1),\\
 v_{32} = v(1, 1, 2, 0, 1, 0),& v_{33} = v(2, 2, 0, 1, 0, 0),& v_{34} = v(1, 1, 1, 0, 1,1), & v_{35} = v(1, 1, 1, 1, 0, 1),\\
 v_{36} = v(1, 1, 1, 1, 1, 0),& v_{37} = v(2, 1, 2, 0, 1, 0), & v_{38} = v(3, 2, 0, 1, 0, 0),& v_{39} = v(2, 1, 1, 0, 1, 1),\\
 v_{40} = v(2, 1, 1, 1,0, 1),& v_{41} = v(2, 1, 1, 1, 1, 0),& v_{42} = v(2, 2, 1, 1, 0, 0),& v_{43} = v(1, 1, 2, 0, 1, 1), \\
 v_{44} = v(2, 2, 0, 1, 0, 1),& v_{45} = v(1, 1, 2, 1, 1, 0),& v_{46} = v(1, 1, 1, 1, 1, 1),& v_{47} = v(3, 2, 1, 1, 0, 0),\\
 v_{48} = v(2, 1, 2, 0, 1, 1),& v_{49} = v(3, 2, 0, 1, 0, 1),& v_{50} = v(2, 1, 2, 1, 1, 0),& v_{51} = v(2, 1, 1, 1, 1, 1),\\
 v_{52} = v(2, 2,1, 1, 0, 1),& v_{53} = v(2, 2, 1, 1, 1, 0), & v_{54} = v(1, 1, 2, 1, 1, 1),& v_{55} = v(3, 2, 1, 1, 0, 1),\\
 v_{56} = v(3, 2, 1, 1, 1, 0),& v_{57} = v(2, 1, 2, 1, 1, 1),& v_{58} = v(2,2, 2, 1, 1, 0),& v_{59} = v(2, 2, 1, 1, 1, 1),\\
 v_{60} = v(3, 2, 2, 1, 1, 0),& v_{61} = v(3, 2, 1, 1, 1, 1),& v_{62} = v(2, 2, 2, 1, 1, 1),& v_{63} = v(3, 2, 2, 1, 1,1).\\
    \end{array}
\end{equation}

In order to avoid irrationalities (denominators containing roots) in our computations, we made a rational rescaling of operators of the representation  with the diagonal matrix $M_4$: $\Tilde{E}_{1,2} = M_4 E_{1,2} M_4^{-1}$, $\Tilde{F}_{1,2} = M_4 F_{1,2} M_4^{-1}$, $\Tilde{K}_{1,2} = K_{1,2}$, where the matrix $M_4$ is
{\small
\begin{equation}
    \begin{array}{ll}
     \text{ } & M_4\, = \, \\
        \text{ } & {\rm diag} \left( \sqrt{[\mu_2 +1]} \sqrt{[\mu_1 +1]} \sqrt{[\mu_2 +\mu_1 +2]},\sqrt{[\mu_2 +1]} \sqrt{[\mu_1 +1]} \sqrt{[\mu_2 +\mu_1 +2]},
        \sqrt{[\mu_2 +1]} \sqrt{[\mu_1 +2]} \sqrt{[\mu_2+\mu_1 +2]},\right.
        \\ &
        \sqrt{[\mu_2 +2]} \sqrt{[\mu_1 +1]} \sqrt{[\mu_2 +\mu_1 +3]},\sqrt{[\mu_2 +1]} \sqrt{[\mu_1 +2]} \sqrt{[\mu_2 +\mu_1 +2]},\sqrt{[\mu_2 +2]} \sqrt{[\mu_1 +1]}
   \sqrt{[\mu_2 +\mu_1 +3]},
   \\ &
   \sqrt{[\mu_2 +1]} \sqrt{[\mu_1 ]} \sqrt{[\mu_2 +\mu_1 +2]},\sqrt{[\mu_2 +2]} \sqrt{[\mu_1 +2]} \sqrt{[\mu_2 +\mu_1 +3]},\sqrt{[\mu_2 ]} \sqrt{[\mu_1
   +1]} \sqrt{[\mu_2 +\mu_1 +2]},
   \\ &
   \sqrt{[\mu_2 +1]} \sqrt{[\mu_1 ]} \sqrt{[\mu_2 +\mu_1 +2]},\sqrt{[\mu_2 +2]} \sqrt{[\mu_1 +2]} \sqrt{[\mu_2 +\mu_1 +3]},\sqrt{[\mu_2 ]}
   \sqrt{[\mu_1 +1]} \sqrt{[\mu_2 +\mu_1 +2]},
   \\ &
   \sqrt{[\mu_2 +1]} \sqrt{[\mu_1 +1]} \sqrt{[\mu_2 +\mu_1 +2]},\sqrt{[\mu_2 +2]} \sqrt{[\mu_1 ]} \sqrt{[\mu_2 +\mu_1 +3]},\sqrt{[\mu_2]} \sqrt{[\mu_1 +1]} \sqrt{[\mu_1 +2]} \sqrt{[\mu_1 +3]} \sqrt{[\mu_2 +\mu_1 +2]},
   \\ &
   \sqrt{[\mu_2 +1]} \sqrt{[\mu_1 +1]} \sqrt{[\mu_2 +\mu_1 +1]} [\mu_2 +\mu_1 +2],\sqrt{[\mu_2 +1]}
   \sqrt{[\mu_1 +1]} \sqrt{[\mu_2 +\mu_1 +3]},\sqrt{[\mu_2 +1]} \sqrt{[\mu_1 +1]} \sqrt{[\mu_2 +\mu_1 +2]},\\&\sqrt{[\mu_2 +2]} \sqrt{[\mu_1 ]} \sqrt{[\mu_2 +\mu_1 +3]},\sqrt{[\mu_2
   ]} \sqrt{[\mu_1 +1]} \sqrt{[\mu_1 +2]} \sqrt{[\mu_1 +3]} \sqrt{[\mu_2 +\mu_1 +2]},
   \\ &
   \sqrt{[\mu_2 +1]} \sqrt{[\mu_1 +1]} \sqrt{[\mu_2 +\mu_1 +1]} [\mu_2 +\mu_1 +2],\sqrt{[\mu_2 +1]}
   \sqrt{[\mu_1 +1]} \sqrt{[\mu_2 +\mu_1 +3]},\sqrt{[\mu_2 +2]} \sqrt{[\mu_1 +1]} \sqrt{[\mu_2 +\mu_1 +3]},
   \\ &
   \sqrt{[\mu_2 ]} \sqrt{[\mu_1 +2]} \sqrt{[\mu_2 +\mu_1 +2]},\sqrt{[\mu_2+1]} \sqrt{[\mu_1 ]} \sqrt{[\mu_2 +\mu_1 +1]},\sqrt{[\mu_2 +1]} \sqrt{[\mu_1 +1]} \sqrt{[\mu_1 +2]} \sqrt{[\mu_1 +3]} \sqrt{[\mu_2 +\mu_1 +3]},\\&\sqrt{[\mu_2 +2]} \sqrt{[\mu_1 +1]}
   \sqrt{[\mu_2 +\mu_1 +2]},\sqrt{[\mu_2 +2]} \sqrt{[\mu_1 +1]} \sqrt{[\mu_2 +\mu_1 +3]},\sqrt{[\mu_2 ]} \sqrt{[\mu_1 +2]} \sqrt{[\mu_2 +\mu_1 +2]},
   \\ &
   \sqrt{[\mu_2 +1]} \sqrt{[\mu_1]} \sqrt{[\mu_2 +\mu_1 +1]},\sqrt{[\mu_2 +1]} \sqrt{[\mu_1 +1]} \sqrt{[\mu_1 +2]} \sqrt{[\mu_1 +3]} \sqrt{[\mu_2 +\mu_1 +3]},\\&\sqrt{[\mu_2 +2]} \sqrt{[\mu_1 +1]} \sqrt{[\mu_2+\mu_1 +2]},\sqrt{[\mu_2 ]} \sqrt{[\mu_1 +1]} \sqrt{[\mu_2 +\mu_1 +2]},\sqrt{[\mu_2 +1]} \sqrt{[\mu_1 -1]} \sqrt{[\mu_1 ]} \sqrt{[\mu_1 +1]} \sqrt{[\mu_2 +\mu_1+1]},
   \\ &
   \sqrt{[\mu_2+1]} \sqrt{[\mu_1 +2]} \sqrt{[\mu_2 +\mu_1 +3]},\sqrt{[\mu_2 +2]} \sqrt{[\mu_1 ]} \sqrt{[\mu_2 +\mu_1 +2]},\sqrt{[\mu_2 ]} \sqrt{[\mu_1 +1]} \sqrt{[\mu_2 +\mu_1+1]},
   \\ &
   \sqrt{[\mu_2 ]} \sqrt{[\mu_1 +1]} \sqrt{[\mu_2 +\mu_1 +2]},\sqrt{[\mu_2 +1]} \sqrt{[\mu_1 -1]} \sqrt{[\mu_1 ]} \sqrt{[\mu_1 +1]} \sqrt{[\mu_2 +\mu_1 +1]},\sqrt{[\mu_2 +1]}
   \sqrt{[\mu_1 +2]} \sqrt{[\mu_2 +\mu_1 +3]},
   \\ &
   \sqrt{[\mu_2 +2]} \sqrt{[\mu_1 ]} \sqrt{[\mu_2 +\mu_1 +2]},\sqrt{[\mu_2 ]} \sqrt{[\mu_1 +1]} \sqrt{[\mu_2 +\mu_1 +1]},\sqrt{[\mu_2+1]} \sqrt{[\mu_1 +1]} \sqrt{[\mu_2 +\mu_1 +1]},
   \\ &
   \sqrt{[\mu_2 +1]} \sqrt{[\mu_1 +1]} \sqrt{[\mu_2 +\mu_1 +3]},\sqrt{[\mu_2 +2]} \sqrt{[\mu_1 -1]} \sqrt{[\mu_1 ]} \sqrt{[\mu_1 +1]}
   \sqrt{[\mu_2 +\mu_1 +2]},\sqrt{[\mu_2 ]} \sqrt{[\mu_1 +2]} \sqrt{[\mu_2 +\mu_1 +1]},
   \\ &
   \sqrt{[\mu_2 +1]} \sqrt{[\mu_1 +1]} \sqrt{[\mu_2 +\mu_1 +2]},\sqrt{[\mu_2 +1]} \sqrt{[\mu_1+1]} \sqrt{[\mu_2 +\mu_1 +1]},\sqrt{[\mu_2 +1]} \sqrt{[\mu_1 +1]} \sqrt{[\mu_2 +\mu_1 +3]},
   \\ &
   \sqrt{[\mu_2 +2]} \sqrt{[\mu_1 -1]} \sqrt{[\mu_1 ]} \sqrt{[\mu_1 +1]} \sqrt{[\mu_2+\mu_1 +2]},\sqrt{[\mu_2 ]} \sqrt{[\mu_1 +2]} \sqrt{[\mu_2 +\mu_1 +1]},\sqrt{[\mu_2 +1]} \sqrt{[\mu_1 +1]} \sqrt{[\mu_2 +\mu_1 +2]},
   \\ &
   \sqrt{[\mu_2 +2]} \sqrt{[\mu_1 +1]}\sqrt{[\mu_2 +\mu_1 +2]},\sqrt{[\mu_2 ]} \sqrt{[\mu_1 ]} \sqrt{[\mu_2 +\mu_1 +1]},\sqrt{[\mu_2 +1]} \sqrt{[\mu_1 +2]} \sqrt{[\mu_2 +\mu_1 +2]},
   \\ &
   \sqrt{[\mu_2 +2]} \sqrt{[\mu_1+1]} \sqrt{[\mu_2 +\mu_1 +2)},\sqrt{[\mu_2 )} \sqrt{[\mu_1 )} \sqrt{[\mu_2 +\mu_1 +1)},\sqrt{[\mu_2 +1)} \sqrt{[\mu_1 +2)} \sqrt{[\mu_2 +\mu_1 +2)},
   \\ &
   \sqrt{[\mu_2 )} \sqrt{[\mu_1+1]} \sqrt{[\mu_2 +\mu_1 +1]},\sqrt{[\mu_2 +1]} \sqrt{[\mu_1 ]} \sqrt{[\mu_2 +\mu_1 +2]},\sqrt{[\mu_2 ]} \sqrt{[\mu_1 +1]} \sqrt{[\mu_2 +\mu_1 +1]},
   \\ &
   \sqrt{[\mu_2 +1]}\sqrt{[\mu_1 ]} \sqrt{[\mu_2 +\mu_1 +2]},\sqrt{[\mu_2 +1]} \sqrt{[\mu_1 +1]} \sqrt{[\mu_2 +\mu_1 +2]},  \left. \sqrt{[\mu_2 +1]} \sqrt{[\mu_1 +1]} \sqrt{[\mu_2 +\mu_1 +2]}  \right ).
    \end{array}
\end{equation}}

\subsection{\R-matrices in representations $W_{m,3}(q^{\mu_1},q^{\mu_2})$ \label{Rsl3}}
In this section, we discuss how to calculate the \R-matrix in  representations $W_{m,3}(q^{\mu_1},q^{\mu_2}) = W_{m,3}(\l_1,\l_2)$ of $\U_q(sl_3)$.
In the case of \uqslthree{}, the Cartan matrix is $a_{ij} = \begin{pmatrix} 2 & -1\\ -1 & 2 \end{pmatrix}$, and the positive roots $\Phi^{+}$ are
\begin{equation}
    \begin{array}{lll}
    E_{\beta_1} =  E_1, \,\,\,\,\,\,\,\,\,\, &  E_{\beta_2}   = q^{-1}E_2 E_1 - E_1 E_2, \,\,\,\,\,\,\,\,\,\, &  E_{\beta_3}  = E_2,\\
        F_{\beta_1} = F_1, &F_{\beta_2} = qF_1 F_2 - F_2 F_1, & F_{\beta_3} = F_2 \\
    \end{array}
\end{equation}
so that formula (\ref{UniR}) takes the form
\begin{equation}
\begin{split}
\left.\mathcal{R}_{u}\right |_{sl_3}  = &  P\, q^{\frac{2}{3}h_1 \otimes h_1 + \frac{1}{3} h_1 \otimes h_2 + \frac{1}{3} h_2 \otimes h_1 + \frac{2}{3} h_2 \otimes h_2}
 \\
&  {\rm exp}_q ((q-q^{-1}) \text{ } E_1 \otimes F_1) \text{ } {\rm exp}_q \text{ } ((q-q^{-1}) E_{12} \otimes F_{12}) \text{ } {\rm exp}_q \text{ } ((q-q^{-1}) E_2 \otimes F_2)
\end{split}
\end{equation}

In order to work in the topological framing, we normalize the \R-matrices
\begin{equation}
     \r_{m,3} =  q^{-2/3\left(\mu_1^2+\mu_1 \mu_2 +\mu_2^2 \right)} \l_1^{2m-2} \l_2^{2m-2}  \left. \r_u\right|_{sl_3},
\end{equation}
or, when we evaluating the \R-matrix that acts on different representations,
\begin{equation}
     \r_{m,3}(\l^{(1)}_{i},\l^{(2)}_{i}) =  q^{-2/3\left(\m_1^{(1)}\m_1^{(2)}+ \frac{1}{2} \left(\m_1^{(1)}\mu_2^{(2)} + \mu_1^{(2)}\mu_2^{(1)} \right)    + \m_2^{(1)}\m_2^{(2)} \right)} \left( \l_1^{(1)}  \l_1^{(2)}  \l_2^{(1)}  \l_2^{(2)} \right)^{m-1}\left. \r_u\right|_{sl_3},
\end{equation}

and the weight matrix is
\begin{equation}
    \begin{array}{l}
        \W_{m,3} = \l_1^{-2m}\l_2^{-2m} K_1^{2} K_2^{2},\\
    \end{array}
\end{equation}
where $K_1$ and $K_2$ and the operators that enter $\left. R_u\right|_{sl_3}$ belong to the corresponding nilpotent representation  with parameters $W_{m,3}(q^{\mu_1},q^{\mu_2})$, $q^{\mu_1} = \l_1$, $q^{\mu_2} = \l_2$.

We also calculated eigenvalues of these \R-matrices. $\r_{2,3}$ has 8 eigenvalues, each of them corresponds to a subspace of dimension 8:
\begin{equation}
    \begin{array}{cccc}
        q^{\pm 1}, \,\,\,\,\, & \left(\l_1^{(1)}\l_1^{(2)} \right)^{\pm 1}, \,\,\,\,\, &  \left(\l_2^{(1)}\l_2^{(2)} \right)^{\pm 1}, \,\,\,\,\, & -\left(\l_1^{(1)}\l_1^{(2)} \l_2^{(1)}\l_2^{(2)} \right)^{\pm 1}.
    \end{array}
\end{equation}
$R_{3,3}$ has 27 eigenvalues, each of them corresponds to a subspace of dimension 27. Here are 24 of them (there are 3 more missing for computational reasons):

{\small
\begin{equation}
    \begin{array}{lllll}
        \left(\l_1^{(1)}\l_1^{(2)} \l_2^{(1)}\l_2^{(2)} \right)^{2}, \,\,\,\,\, &   -\l_1^{(1)}\l_1^{(2)}\left(\l_2^{(1)}\l_2^{(2)} \right)^{2},\,\,\,\,\,  &   q^{2}\left(\l_1^{(1)}\l_1^{(2)} \right)^{2}, \,\,\,\,\, & \pm \l_2^{(1)}\l_2^{(2)},  \,\,\,\,\,  &   -\l_2^{(1)}\l_2^{(2)}\left(\l_1^{(1)}\l_1^{(2)}\right)^{-1}  \\
       q^2\left(\l_1^{(1)}\l_1^{(2)} \l_2^{(1)}\l_2^{(2)} \right)^{-2},   \,\,\,   &-\l_2^{(1)}\l_2^{(2)}\left(\l_1^{(1)}\l_1^{(2)} \right)^{2}, \,\,\,\,\, & q^2 \left(\l_2^{(1)}\l_2^{(2)} \right)^{2},  \,\,\,\,\, & \pm\l_1^{(1)}\l_1^{(2)},  \,\,\,\,\,&-\l_1^{(1)}\l_1^{(2)}\left(\l_2^{(1)}\l_2^{(2)}\right)^{-1} \\
       & -\left(\l_2^{(1)}\l_2^{(2)}\right)^{-1}\left(\l_1^{(1)}\l_1^{(2)} \right)^{-2}, \,\,\, &\left(\l_1^{(1)}\l_1^{(2)} \right)^{-2},  \,\,\,&\pm q^2 \left(\l_2^{(1)}\l_2^{(2)}\right)^{-1} \,\,\,&\\
       &-\left(\l_2^{(1)}\l_2^{(2)}\right)^{-2}\left(\l_1^{(1)}\l_1^{(2)} \right)^{-1}, \,\,\,\,\, & \left(\l_2^{(1)}\l_2^{(2)} \right)^{-2}, \,\,\,\,\, &\pm q^2 \left(\l_1^{(1)}\l_1^{(2)}\right)^{-1} &\\
       &&\pm q^2 \l_1^{(1)}\l_1^{(2)} \l_2^{(1)}\l_2^{(2)},  \,\, &&\\
      &&\pm \left( \l_1^{(1)}\l_1^{(2)} \l_2^{(1)}\l_2^{(2)} \right)^{-1} \,\,&&\\
    \end{array}
\end{equation}
}

\subsection{\R-matrices in representation $W_{2,4}(q^{\mu_1},q^{\mu_2},q^{\mu_3})$ \label{Rsl4}}
In this section, we discuss how to calculate the \R-matrix in representation $W_{2,4}(q^{\mu_1},q^{\mu_2},q^{\mu_3}) = W_{2,4}(\l_1,\l_2,\l_3)$ of $\U_q(sl_4)$. In the case of \uqslfour{}, the Cartan matrix is $a_{ij} = \begin{pmatrix} 2 & -1 & 0 \\ -1 & 2 & -1 \\ 0 & -1 & 2 \end{pmatrix}$, and the positive roots $\Phi^{+}$ are
\begin{equation}
    \begin{array}{l}
        F_{\beta_1} = F_1 , \\
        F_{\beta_2} = q F_1 F_2 - F_2 F_1 ,\\
        F_{\beta_3} = F_2 ,\\
        F_{\beta_4} = q^2 F_1 F_2 F_3 -q F_2 F_3 F_1 - q F_3 F_1 F_2+ F_3 F_2 F_1,\\
        F_{\beta_5} = q F_2 F_3 -F_3 F_2 ,\\
        F_{\beta_6} = F_3 ,\\ \\
        E_{\beta_1} = E_1,\\
        E_{\beta_2} = q^{-1}E_2E_1-E_1E_2,\\
        E_{\beta_3} = E_2,\\
        E_{\beta_4} = E_1E_2E_3-q^{-1}E_1E_23E_2-q^{-1}E_{2}E_1E_3-q^{-2} E_3E_2E_1,\\
        E_{\beta_5} =q^{-1} E_3 E_2-E_2E_3 ,\\
        E_{\beta_6} = E_3,\\
    \end{array}
\end{equation}
In order to avoid additional framing coefficients, we normalise the universal \R-matrix
\begin{equation}
    \r_{2,4} = q^{-\left( \frac{3}{4} \mu_1^2 + \mu_1 \mu_2 +\frac{1}{2} \mu_1 \mu_3 +\mu_2^2+\mu_2 \mu_3 + \frac{3}{4} \mu_3^2 \right)} \l_1^3 \l_2^4 \l_3^3 \left. \r_u \right|_{sl_4}
\end{equation}
or, if working with two different representations,
\begin{eqnarray}
\r_{2,4}\left(q^{\m^{(1)}_{i}},q^{\m^{(2)}_{i}}\right) = q^{-\left(  \frac{3}{4}\m_1^{(1)}\m_1^{(2)}+\frac{1}{2}\left(\m_1^{(1)}\m_2^{(2)}+\m_1^{(2)}\m_2^{(1)} \right)+\frac{1}{4}\left(   \m_1^{(1)} \m_3^{(2)} + \m_1^{(2)}\m_3^{(1)}\right)+\m_2^{(1)}\m_2^{(2)}+\frac{1}{2}\left(   \m_2^{(1)} \m_3^{(2)} + \m_2^{(2)}\m_3^{(1)}\right)+\frac{3}{4}\m_3^{(1)}\m_3^{(2)} \right)} \nn \\ q^{\frac{3}{2}\left(\m_1^{(1)}+\m_1^{(2)} \right)} q^{2\left(\m_2^{(1)}+\m_2^{(2)} \right)}q^{\frac{3}{2}\left(\m_3^{(1)}+\m_3^{(2)} \right)}  \left. \r_u \right|_{sl_4},\,\,\,\,\,\,\,\,\,\,
\end{eqnarray}
and the weight matrix is
\begin{equation}
    \mathcal{W}_{2,4} = \lambda_1^{-6}\l_2^{-8}\l_3^{-6} K_1^3 K_2^4 K_3^3.
\end{equation}

\subsection{Invariants of knots and links}
The reduced polynomials of knots and links can be calculated by formula (\ref{ppolynomial}) with $\r_{m,3}$, $\r_{m,4}$ and $\W_{m,3}$, $\W_{m,4}$. In order to get invariants of knots and links, one should use normalization coefficient (\ref{norm}):
\begin{equation}
   \begin{array}{l}
        \Xi_{m,3}(\l) = \xi_m(\lambda_1)\, \xi_m(\l_2)\, \xi_m(q\lambda_1 \l_2),\\
\Xi_{m,4}(\l) = \xi_m(\lambda_1)\, \xi_m(\l_2)\, \xi_m(\l_3)\,\xi_m(q\lambda_1\l_2)\,\xi_m(q\l_2\l_3)\,\xi_m(q^2\lambda_1\l_2\l_3)
   \end{array} \end{equation}
Then, the invariants $\P_{m,N}^{\L}(\l^{(k)}_i)$ (where $1 \leq i \leq N-1$,  $k$ enumerates the components of link, $\l^{(1)}_i$ is color of the open component) of knots and links at roots of unity are
\begin{equation}
\P_{m,N}^{\L}(\l^{(k)}_i) = {P_{m,N}^{\L}(\l^{(k)}_i) \over \Xi_{m,N}(\l^{(1)}_i)}.
\end{equation}

\subsection{Connection with HOMFLY-PT and Alexander polynomials}
The polynomials $P_{m,N}^{\L}(\l^{(k)}_i)$ and the invariants are also connected with the HOMFLY-PT and Alexander polynomials from not-a-root-of-unity case. In the case of \uqslthree{}:

\begin{equation}
     P^{\K}_{2,3}\,(\lambda_1, \l_2=1) = \A^{\K} (\lambda^{2}_1),
\end{equation}
\begin{equation}
     P^{\K}_{3,3}\,(q,\lambda_1, \l_2=1) =\A^{\K} (\lambda_1) \A^{\K} (\lambda^{3}_1),
\end{equation}
These expressions are the same for $\lambda_1 = 1$ because $P_{m,3}^{\K}\,(\lambda_1,\l_2) = P_{m,3}^{\K}\,(\l_2,\lambda_1)$.
In the case of \,\uqslfour{}, the symmetry between the parameters is more complicated, and there are the following connections between $ P_{2,4}^{\K}(q,\l_1,\l_2,\l_3)$ and the Alexander polynomials.
\begin{equation}
    P_{2,4}^{\K}(q,\l_1=1,\l_2,\l_3=1) = \A^{\K}(\l_2^2),
\end{equation}
\begin{equation}
P_{2,4}^{\K}(q,\l_1=1,\l_2=1,\l_3)  = \A^{\K}(\l_3)\A(\l_3^2),
\end{equation}
\begin{equation}
    P_{2,4}^{\K}(q,\l_1,\l_2=1,\l_3=1)  = \A^{\K}(\l_1)\A(\l_1^2).
\end{equation}

The invariants of links $\P_{2,3}^{\L}$ are also connected with the Alexander polynomials of links $\D^{\L}$, however the connection is more elaborate:
\begin{equation}
    \P_{2,3}^{L_2 a_1} (q,\l_1^{(1)},\l_2^{(1)},\l_1^{(2)},\l_2^{(2)}) = q \, \D^{L_2a_1},
\end{equation}
\begin{equation}
    \P_{2,3}^{L_5 a_1}(q,\l_1^{(1)},\l_2^{(1)} = q,\l_1^{(2)},\l_2^{(2)} = q) = -4q \left(\left(\D^{L_5 a_1}(\l_1^{(1)},\l_1^{(2)} )\right)^2-1\right),\\
\end{equation}
\begin{equation}
    \P_{2,3}^{L_6 a_4}(q,\l_1^{(1)},\l_2^{(1)} = q,\l_1^{(2)},\l_2^{(2)} = q,\l_1^{(2)},\l_2^{(2)} = q) = 8q \left(\left(\D^{L_6 a_4}(\l_1^{(1)},\l_1^{(2)},\l_1^{(3)} )\right)^2-2\right).\\
\end{equation}

\section{Conclusion}

The knot/link invariants associated with representations of $\mathcal{U}_q(sl_N)$ are a part of modern theoretical physics. In this paper, we considered the invariants that emerge when the parameter of quantization $q$ is a root of unity. This case is special because the representation structure of $\mathcal{U}_q(sl_N)$ drastically changes at roots of unity, and new types of representations emerge. The most interesting ones are the nilpotent representations with parameters, which produce a new multi-parametric family of invariants $\P_{m,N}^{\L}(\l^{(k)}_i)$. The corresponding invariants of knots $\P_{m,N}^{\K}(\l_i)$ depend on $N-1$ parameters, and the invariants of links $\P_{m,N}^{\L}(\l^{(k)}_i)$ depend on $l\, (N-1)$ parameters, where $l$ is the number of link components.

To determine the invariants of knots and links at roots of unity, we used a modified Reshetikhin-Turaev method to find polynomials of tangles and then restore the normalization coefficient to get invariants.

Though the resulting polynomials and invariants, at some peculiar point, coincide with the HOMFLY-PT polynomials, and are connected with the Alexander polynomials, they generically provide a new type of invariants of knots and links.

We considered all possible regular series of irreducible finite-dimensional representations of $\mathcal{U}_q(sl_N)$ at $q$ equal to a root of unity, and hence completed the classification of all invariants associated with $\mathcal{U}_q(sl_N)$. However, we did not deal with exotic types of representations: atypical and partially periodic, it would be interesting to check if they can give rise to new knot/link invariants.
It would be also interesting to construct invariants for other classical groups and check if the existing relations of such invariants at $|q|<1$ with the HOMFLY-PT polynomials (see, e.g., the relation between the Kauffman and HOMFLY-PT polynomials \cite{Kauffman}) still persists at $q$ equal to a root of unity.

\section*{Acknowledgements}

This work was partly supported by Russian Science Foundation grant No 18-71-10073..

\section{Appendix A. Polynomials $P_{m,N}^{\K}$ of knots \label{appA}}
\subsection{\uqsltwo{}}

\paragraph{Trefoil}
\begin{equation}
    \begin{array}{ll}
     P_{2,2}^{3_1}(\l)\,\,\,\, = & \l^2-1+\l^{-2} , \\
       P_{3,2}^{3_1}(\l,q) = & q^{-1}\l^{-4}(\lambda ^6-2 \lambda ^4+(\lambda ^8-\lambda ^6+\lambda ^4+\lambda ^2-1) q+1), \\
        P_{4,2}^{3_1}(\l,q) =&  -q^{-6}\l^{-6}(\lambda ^{10}-\lambda ^8-2 \lambda ^6+\lambda ^4+\lambda ^2+(\lambda ^{12}-\lambda ^8+\lambda ^6+\lambda ^4-1) q^2 )   ,\\
         P_{5,2}^{3_1}(\l,q) = &  -q^{-8}\l^{-8} (\lambda ^{12}-\lambda ^{10}-2 \lambda ^8+\lambda ^6+\lambda ^2+(\lambda ^{14}-\lambda ^{10}+\lambda ^8+\lambda ^6+\lambda ^4-\lambda ^2-1) \lambda ^2 q^3+  \\
         & (\lambda ^{12}-2 \lambda ^{10}+\lambda ^6+\lambda ^2-1) q^2+(\lambda
   ^{14}-2  \lambda ^{12}+\lambda ^8+\lambda ^4-\lambda ^2) q )  ,\\
          P_{6,2}^{3_1}(\l,q) = & q^{-10}\lambda^{-10}(\lambda ^2 (\lambda ^{18}-\lambda ^{14}+2 \lambda ^{10}+3 \lambda ^8-2 \lambda ^4-\lambda ^2-1) + \\
          & (-\lambda ^{20}-\lambda ^{18}+\lambda ^{16}+2 \lambda ^{14}-3 \lambda ^{10}-2 \lambda ^8+\lambda ^4+\lambda ^2+1) q^2 )   ,\\
           P_{7,2}^{3_1}(\l,q) = &-q^{-12}\l^{-12}  (\lambda ^4 (\lambda ^{14}-\lambda ^{12}-\lambda ^{10}-2 \lambda ^8+\lambda ^6+\lambda ^4+1)-(-\lambda ^{24}+\lambda ^{18}-\lambda ^{16}-2 \lambda ^{14}-\lambda ^{10}+\lambda ^8+\lambda ^6+\lambda ^4) q^5+ \\
          &  (\lambda ^{18}-2 \lambda
   ^{16}+\lambda ^{12}+\lambda ^4-\lambda ^2) q^4+\lambda ^4 (\lambda ^{18}-\lambda ^{16}-2 \lambda ^{14}+\lambda ^{12}+\lambda ^{10}+\lambda ^8+\lambda ^4-\lambda ^2-1) q^3+ \\
          &  (\lambda ^4-1)^2 (\lambda ^{10}-\lambda
   ^8-\lambda ^4-1) q^2+\lambda ^4 (\lambda ^{16}-2 \lambda ^{14}+\lambda ^{10}+\lambda ^6+\lambda ^2-2) q ).\\
    \end{array}
\end{equation}
\paragraph{Figure-eight knot}
\begin{equation}
    \begin{array}{ll}
           P_{2,2}^{4_1}(\l)\,\,\,\, = & -\l^2+3-\l^{-2} , \\
           P_{3,2}^{4_1}(\l,q) = & q^{-2}\l^{-4}(\lambda ^8-5 \lambda ^4+3 \lambda ^2+(-3 \lambda ^6+5 \lambda ^4-1) q ), \\
            P_{4,2}^{4_1}(\l,q) = & \l^{-6}(\lambda ^2 (-3 \lambda ^8+7 \lambda ^4-3)-(\lambda ^{12}-6 \lambda ^8+6 \lambda ^4-1) q^2), \\
             P_{5,2}^{4_1}(\l,q) = &-\l^{-8}(\lambda ^2 (6 \lambda ^{10}-10 \lambda ^6-\lambda ^4+\lambda ^2+3)+\lambda ^2 (\lambda ^{14}-6 \lambda ^{10}-\lambda ^8+9 \lambda ^4-3) q^3 + \\ &  (5 \lambda ^{12}-8 \lambda ^{10}+\lambda ^4+3 \lambda ^2-1) q^2+\lambda ^2 (3
   \lambda ^{12}-5 \lambda ^{10}-\lambda ^8+\lambda ^4+5 \lambda ^2-3) q ), \\
              P_{6,2}^{4_1}(\l,q) = & -q^{-8}\l^{-10}(\lambda ^{20}+3 \lambda ^{18}-12 \lambda ^{14}-14 \lambda ^{12}+14 \lambda ^8+12 \lambda ^6-3 \lambda ^2  +\\& (-3 \lambda ^{18}-7 \lambda ^{16}+14 \lambda ^{12}+15 \lambda ^{10}-12 \lambda ^6-7 \lambda ^4+1) q^2-1 ), \\
               P_{7,2}^{4_1}(\l,q) = & q^{-6}\l^{-12}((6 \lambda ^{16}+3 \lambda ^{14}-3 \lambda ^{10}-19 \lambda ^8-2 \lambda ^6+2 \lambda ^4+11 \lambda ^2+1) \lambda ^4 - \\& (\lambda ^{16}+11 \lambda ^{14}+2 \lambda ^{12}-2 \lambda ^{10}-19 \lambda ^8-3 \lambda ^6+3 \lambda ^2+6) \lambda
   ^4 q^5+\\& (\lambda ^{18}-2 \lambda ^{10}-2 \lambda ^8-16 \lambda ^6+14 \lambda ^4+2 \lambda ^2+2) \lambda ^6 q^4- \\& (3 \lambda ^{16}+13 \lambda ^{14}-18 \lambda ^{10}-3 \lambda ^8-2 \lambda ^6+2 \lambda ^4+\lambda ^2+3) \lambda ^2
   q^3+\\& (3 \lambda ^{16}+\lambda ^{14}+2 \lambda ^{12}-2 \lambda ^{10}-3 \lambda ^8-18 \lambda ^6+13 \lambda ^2+3) \lambda ^6 q^2+ \\ & (-2 \lambda ^{18}-2 \lambda ^{16}-14 \lambda ^{14}+16  \lambda ^{12}+2 \lambda ^{10}+2 \lambda ^8-1) q ). \\
    \end{array}
\end{equation}

\paragraph{Knot $5_1$}

\begin{equation}
    \begin{array}{ll}
        P_{2,2}^{5_1}(\l) = & q^4-q^2+1-q^{-2}+q^{-4}, \\
         P_{3,2}^{5_1}(\l,q) = & q^{-1}\l^{-8} (\lambda ^{10}+\lambda ^{16} q-\lambda ^{14} q^2+\lambda ^{12} (q-2)-\lambda ^8 q+\lambda ^6 q^2+\lambda ^4 (q+1)-\lambda ^2 q^2-1 ), \\
          P_{4,2}^{5_1}(\l,q) = & q^{2}\l^{-12} (-\lambda ^{22}+\lambda ^{20}+\lambda ^{18}-\lambda ^{16}+\lambda ^{14}-\lambda ^{10}-\lambda ^8-\lambda ^6+\lambda ^4+\lambda ^2+ \\ &  (-\lambda ^{24}+\lambda ^{20}-\lambda ^{18}+\lambda ^{14}-\lambda ^{12}-\lambda ^{10}+\lambda ^6+\lambda ^4-1) q^2 ), \\
           P_{5,2}^{5_1}(\l,q) = & -q^{2}\l^{-16} ( (\lambda ^{24}-\lambda ^{22}-\lambda ^{20}+\lambda ^{18}-\lambda ^{16}+\lambda ^{12}+\lambda ^{10}+\lambda ^8-\lambda ^6+\lambda ^4-\lambda ^2-1) \lambda ^4+ \\ &  (\lambda ^{32}-\lambda ^{28}+\lambda ^{26}+\lambda ^{18}-2 \lambda ^{14}-\lambda
   ^8+\lambda ^6+\lambda ^2) q^3+\\ & (\lambda ^{28}-2 \lambda ^{26}+\lambda ^{24}-\lambda ^{20}+\lambda ^{18}+\lambda ^{14}-\lambda ^{10}+\lambda ^8-2 \lambda ^6+\lambda ^4) q^2+\\ &  (\lambda ^{30}-2 \lambda ^{28}+\lambda ^{26}-\lambda
   ^{22}+\lambda ^{20}+\lambda ^{16}-\lambda ^{14}-\lambda ^{12}+\lambda ^{10}-2 \lambda ^8+\lambda ^6+1) q), \\
            P_{6,2}^{5_1}(\l,q) = & q^{2}\l^{-20} (\lambda ^{40}-\lambda ^{36}+\lambda ^{32}+\lambda ^{30}+\lambda ^{26}-2 \lambda ^{22}-3 \lambda ^{20}+2 \lambda ^{16}+\lambda ^{14}-\lambda ^{10}+\lambda ^8+\lambda ^6-\\ & (\lambda ^{40}+\lambda ^{38}-\lambda ^{36}-\lambda ^{34}+\lambda ^{26}+2 \lambda
   ^{24}-3 \lambda ^{20}-2 \lambda ^{18}+\lambda ^{14}-\lambda ^{10}+\lambda ^6+\lambda ^4+\lambda ^2) q^2-1 ), \\
             P_{7,2}^{5_1}(\l,q) = & q^{2}\l^{-24} ( \lambda ^{48} (-q^5)+\lambda ^{22} (-q^5+q+1)-\lambda ^6 (q^5+1)+q^5-q^4-\lambda ^{46} q^3-\lambda ^{18} q^3 (q+1)+\\ &  q^3+\lambda ^{44} q (q^2-1)+\lambda ^{36} (-q^2+q+1)+\lambda ^{16} q
   (q^2-q-1)-\lambda ^8 (q^2-1)-q^2+\lambda ^{32} q^3 (q^2-q-1)+\\ & \lambda ^{24} (q^3+q^2+1)+\lambda ^4 q^3 (q^2-1)+\lambda ^{26} (q^5+q^4+q^2)+\lambda ^{28} (q^5+q^3-q^2+q-2)+\\ & \lambda
   ^{38} (-q^4+q^3+q^2)+\lambda ^{34} q (-2 q^4+q^3-q^2+q-1)+\lambda ^{12} (2 q^4-q^3+q^2-q+1)+\\ & \lambda ^{10} (-q^4+q^3+q^2)+\lambda ^{42} (q^5-q^4+q^3-q^2+2 q-1)-\lambda ^{40} (q^5-2
   q^4+q^3-q^2+q-1)+\\ & \lambda ^{20} (-2 q^5+q^4-2 q^3+q^2-q+1)+\lambda ^{14} (q^5-q^4+q^3-q^2+2 q-2)-\lambda ^{30} q (q+1)-\lambda ^2 q+q-1 . \\
    \end{array}
\end{equation}

\subsection{\uqslthree{}}
\paragraph{Trefoil}
\begin{equation}
    \begin{array}{ll}
        P_{2,3}^{3_1} (\l_1,\l_2) = & ( (\lambda _1^8-\lambda _1^6+\lambda _1^4) \lambda _2^8+(-\lambda _1^8+2 \lambda _1^6-2
   \lambda _1^4+\lambda _1^2) \lambda _2^6+(\lambda _1^8-2 \lambda _1^6+\lambda _1^4-2 \lambda
   _1^2+1) \lambda _2^4+\\ & (\lambda _1^6-2 \lambda _1^4+2 \lambda _1^2-1) \lambda
   _2^2+\lambda _1^4-\lambda _1^2+1)   ,\\
   \end{array}
   \end{equation}
  {\small \begin{equation}
    \begin{array}{ll}
           P_{3,3}^{3_1} (\l_1,\l_2,q) = & q^{-2}\l_1^{-8}\l_2^{-8}  ( \lambda _1^6+\lambda _1^8 \lambda _2^{16} (\lambda _1^6+\lambda _1^8 q^2-\lambda _1^4
   (q+1)+\lambda _1^2 q^2+1)+\lambda _1^6 \lambda _2^{14} (\lambda _1^{10}-2 \lambda _1^6+3
   \lambda _1^4- \\ &   (2 \lambda _1^8-3 \lambda _1^6+\lambda _1^4+2 \lambda _1^2-1)
   q-1)-\lambda _1^4 \lambda _2^{12} (\lambda _1^{12}+2 \lambda _1^{10}-5 \lambda _1^8+2
   \lambda _1^6+3 \lambda _1^4-3 \lambda _1^2+ \\ & (\lambda _1^{12}-3 \lambda _1^{10}+2 \lambda _1^8+2
   \lambda _1^6-5 \lambda _1^4+\lambda _1^2+1) q+1)+\lambda _1^2 \lambda _2^{10}
   (\lambda _1^{14} q^2-\lambda _1^{12} (q-3)-2 \lambda _1^{10} (q+1)+\\ &  4 \lambda _1^8 q-4 \lambda
   _1^6 q^2-2 \lambda _1^4 (q+1)+\lambda _1^2 (3 q-2)+1)+\lambda _2^8 (\lambda _1^{16}-3
   \lambda _1^{12}+4 \lambda _1^{10}-7 \lambda _1^8+5 \lambda _1^4+ \\ &  (-2 \lambda _1^{14}+5 \lambda
   _1^{12}-4 \lambda _1^{10}+\lambda _1^8+4 \lambda _1^6-2 \lambda _1^4-2 \lambda _1^2+1)
   q-1)+\lambda _2^6 (\lambda _1^{14} q^2-\lambda _1^{12} (q-3)- \\ &  2 \lambda _1^{10} (q+1)+4
   \lambda _1^8 q-4 \lambda _1^6 q^2-2 \lambda _1^4 (q+1)+\lambda _1^2 (3 q-2)+1)- \\ &  \lambda _2^4
   (\lambda _1^{12}+2 \lambda _1^{10}-5 \lambda _1^8+2 \lambda _1^6+3 \lambda _1^4-3 \lambda
   _1^2+(\lambda _1^{12}-3 \lambda _1^{10}+2 \lambda _1^8+2 \lambda _1^6-5 \lambda _1^4+\lambda
   _1^2+1) q+1)+\\ &  \lambda _2^2 (\lambda _1^{10}-2 \lambda _1^6+3 \lambda _1^4-(2
   \lambda _1^8-3 \lambda _1^6+\lambda _1^4+2 \lambda _1^2-1) q-1)+\lambda _1^8 q^2-\lambda
   _1^4 (q+1)+\lambda _1^2 q^2+1)  ,\\
   \end{array}
   \end{equation} }
 {\small  \begin{equation}
    \begin{array}{ll}
            P_{4,3}^{3_1} (\l_1,\l_2) = & \l_1^{-12}\l_2^{-12}  ( \lambda _1^{12} (\lambda _1^{12}-\lambda _1^8+\lambda _1^6+\lambda _1^4-1) \lambda
   _2^{24}+\lambda _1^{10} (-2 \lambda _1^{12}+\lambda _1^{10}+3 \lambda _1^8-2 \lambda _1^6-2
   \lambda _1^4+1) \lambda _2^{22}\\ &  -\lambda _1^8 (\lambda _1^{16}-\lambda _1^{14}-3 \lambda
   _1^{12}+3 \lambda _1^{10}+3 \lambda _1^8-3 \lambda _1^6-3 \lambda _1^4+\lambda _1^2+1) \lambda
   _2^{20}+\\ & \lambda _1^6 (\lambda _1^{18}+3 \lambda _1^{16}-3 \lambda _1^{14}-4 \lambda _1^{12}+5
   \lambda _1^{10}+2 \lambda _1^8-4 \lambda _1^6-3 \lambda _1^4+2 \lambda _1^2+2) \lambda
   _2^{18}+\\ & \lambda _1^4 (\lambda _1^{20}-2 \lambda _1^{18}-3 \lambda _1^{16}+5 \lambda _1^{14}-8
   \lambda _1^{10}+5 \lambda _1^6+3 \lambda _1^4-2 \lambda _1^2-1) \lambda _2^{16}+\\ &  (-2 \lambda
   _1^{22}+3 \lambda _1^{20}+2 \lambda _1^{18}-8 \lambda _1^{16}+4 \lambda _1^{14}+10 \lambda _1^{12}-2
   \lambda _1^{10}-5 \lambda _1^8-3 \lambda _1^6+\lambda _1^4+\lambda _1^2) \lambda
   _2^{14}-\\ & (\lambda _1^{24}-3 \lambda _1^{20}+4 \lambda _1^{18}-10 \lambda _1^{14}+5 \lambda
   _1^{12}+10 \lambda _1^{10}-4 \lambda _1^6-3 \lambda _1^4+1) \lambda _2^{12}+\\ &  \lambda _1^2
   (\lambda _1^{20}-\lambda _1^{18}-3 \lambda _1^{16}+5 \lambda _1^{14}-2 \lambda _1^{12}-10 \lambda
   _1^{10}+4 \lambda _1^8+8 \lambda _1^6+2 \lambda _1^4-3 \lambda _1^2-2) \lambda
   _2^{10}+\\ &  (-\lambda _1^{20}+2 \lambda _1^{18}+3 \lambda _1^{16}-5 \lambda _1^{14}+8 \lambda
   _1^{10}-5 \lambda _1^6-3 \lambda _1^4+2 \lambda _1^2+1) \lambda _2^8+\\ & (2 \lambda _1^{18}-2
   \lambda _1^{16}-3 \lambda _1^{14}+4 \lambda _1^{12}+2 \lambda _1^{10}-5 \lambda _1^8-4 \lambda _1^6+3
   \lambda _1^4+3 \lambda _1^2-1) \lambda _2^6-\\ &  (\lambda _1^{16}-\lambda _1^{14}-3 \lambda
   _1^{12}+3 \lambda _1^{10}+3 \lambda _1^8-3 \lambda _1^6-3 \lambda _1^4+\lambda _1^2+1) \lambda
   _2^4+\lambda _1^2 (\lambda _1^{12}-2 \lambda _1^8+2 \lambda _1^6+3 \lambda _1^4-\lambda
   _1^2-2) \lambda _2^2-\\ &  \lambda _1^{12}+\lambda _1^8-\lambda _1^6-\lambda _1^4+q^2 (-\lambda
   _1^{14} (\lambda _1^8-\lambda _1^6-2 \lambda _1^4+\lambda _1^2+1) \lambda _2^{24}+\\ &  \lambda
   _1^{12} (-\lambda _1^{12}+2 \lambda _1^8-2 \lambda _1^6-3 \lambda _1^4+\lambda _1^2+2)
   \lambda _2^{22}+\\ &  \lambda _1^8 (\lambda _1^{16}+2 \lambda _1^{14}-2 \lambda _1^{12}-3 \lambda
   _1^{10}+4 \lambda _1^8+3 \lambda _1^6-2 \lambda _1^4-2 \lambda _1^2+1) \lambda _2^{20}+ \\ &  \lambda
   _1^6 (2 \lambda _1^{18}-2 \lambda _1^{16}-3 \lambda _1^{14}+4 \lambda _1^{12}+2 \lambda _1^{10}-5
   \lambda _1^8-4 \lambda _1^6+3 \lambda _1^4+3 \lambda _1^2-1) \lambda _2^{18}+\\ &  (-\lambda
   _1^{24}-3 \lambda _1^{22}+4 \lambda _1^{20}+2 \lambda _1^{18}-6 \lambda _1^{16}+2 \lambda _1^{14}+6
   \lambda _1^{12}+2 \lambda _1^{10}-4 \lambda _1^8-3 \lambda _1^6+\lambda _1^4) \lambda
   _2^{16}+\\ &  \lambda _1^4 (-\lambda _1^{20}+\lambda _1^{18}+3 \lambda _1^{16}-5 \lambda _1^{14}+2
   \lambda _1^{12}+10 \lambda _1^{10}-4 \lambda _1^8-8 \lambda _1^6-2 \lambda _1^4+3 \lambda
   _1^2+2) \lambda _2^{14}+\\ &  2 \lambda _1^2 (\lambda _1^{20}-\lambda _1^{18}-2 \lambda _1^{16}+3
   \lambda _1^{14}-2 \lambda _1^{12}-5 \lambda _1^{10}+2 \lambda _1^8+3 \lambda _1^6+2 \lambda
   _1^4-\lambda _1^2-1) \lambda _2^{12}+\\ &  (-2 \lambda _1^{20}+3 \lambda _1^{18}+2 \lambda
   _1^{16}-8 \lambda _1^{14}+4 \lambda _1^{12}+10 \lambda _1^{10}-2 \lambda _1^8-5 \lambda _1^6-3 \lambda
   _1^4+\lambda _1^2+1) \lambda _2^{10}+\\ &  (\lambda _1^{20}+3 \lambda _1^{18}-4 \lambda _1^{16}-2
   \lambda _1^{14}+6 \lambda _1^{12}-2 \lambda _1^{10}-6 \lambda _1^8-2 \lambda _1^6+4 \lambda _1^4+3
   \lambda _1^2-1) \lambda _2^8-\\ & (\lambda _1^{18}+3 \lambda _1^{16}-3 \lambda _1^{14}-4 \lambda
   _1^{12}+5 \lambda _1^{10}+2 \lambda _1^8-4 \lambda _1^6-3 \lambda _1^4+2 \lambda _1^2+2) \lambda
   _2^6+\\ &  (\lambda _1^{16}+2 \lambda _1^{14}-2 \lambda _1^{12}-3 \lambda _1^{10}+4 \lambda _1^8+3
   \lambda _1^6-2 \lambda _1^4-2 \lambda _1^2+1) \lambda _2^4+\\ &  (-2 \lambda _1^{12}+\lambda
   _1^{10}+3 \lambda _1^8-2 \lambda _1^6-2 \lambda _1^4+1) \lambda _2^2+\lambda _1^{10}-\lambda
   _1^8-2 \lambda _1^6+  \lambda _1^4+\lambda _1^2)+1  )  \\
    \end{array}
\end{equation}
}
\paragraph{Figure-eight knot}
{\small
\begin{equation}
    \begin{array}{ll}
        P_{2,3}^{4_1}(\l_1,\l_2) = & \l_1^{-4}\l_2^{-4} ((\lambda _1^8-3 \lambda _1^6+\lambda _1^4) \lambda _2^8-3 \lambda _1^2 (\lambda _1^6-4
   \lambda _1^4+4 \lambda _1^2-1) \lambda _2^6+\\&(\lambda _1^8-12 \lambda _1^6+25 \lambda
   _1^4-12 \lambda _1^2+1) \lambda _2^4+ 3 (\lambda _1^6-4 \lambda _1^4+4 \lambda _1^2-1)
   \lambda _2^2+\lambda _1^4-3 \lambda _1^2+1) \\
    \end{array}
\end{equation}
\begin{equation}
    \begin{array}{ll}
        P_{3,3}^{4_1}(\l_1,\l_2) = & q^{-2}\l_1^{-8}\l_2^{-8} (\lambda _1^{10} (\lambda _1^6-5 \lambda _1^2+3) \lambda _2^{16}-3 \lambda _1^6 (4 \lambda
   _1^8-8 \lambda _1^6+4 \lambda _1^2-1) \lambda _2^{14}+\\& \lambda _1^4 (-5 \lambda _1^{12}+24
   \lambda _1^{10}-72 \lambda _1^6+57 \lambda _1^4-5) \lambda _2^{12}+3 \lambda _1^4 (\lambda
   _1^{12}-24 \lambda _1^8+40 \lambda _1^6-24 \lambda _1^2+8) \lambda _2^{10}+\\& (-12 \lambda
   _1^{14}+57 \lambda _1^{12}-163 \lambda _1^8+120 \lambda _1^6-12 \lambda _1^2+1) \lambda _2^8+3
   \lambda _1^2 (\lambda _1^{12}-24 \lambda _1^8+40 \lambda _1^6-24 \lambda _1^2+8) \lambda
   _2^6+\\& (-5 \lambda _1^{12}+24 \lambda _1^{10}-72 \lambda _1^6+57 \lambda _1^4-5) \lambda
   _2^4-3 (4 \lambda _1^8-8 \lambda _1^6+4 \lambda _1^2-1) \lambda _2^2+\lambda _1^2
   (\lambda _1^6-5 \lambda _1^2+3)-\\& (\lambda _1^8 (3 \lambda _1^6-5 \lambda
   _1^4+1) \lambda _2^{16}+3 \lambda _1^8 (\lambda _1^8-4 \lambda _1^6+8 \lambda _1^2-4)
   \lambda _2^{14}+\lambda _1^4 (-5 \lambda _1^{12}+57 \lambda _1^8-72 \lambda _1^6+24 \lambda
   _1^2-5) \lambda _2^{12}+ \\& 3 (8 \lambda _1^{14}-24 \lambda _1^{12}+40 \lambda _1^8-24 \lambda
   _1^6+\lambda _1^2) \lambda _2^{10}+\lambda _1^2 (\lambda _1^{14}-12 \lambda _1^{12}+120
   \lambda _1^8-163 \lambda _1^6+57 \lambda _1^2-12) \lambda _2^8+ \\& 3 (8 \lambda _1^{12}-24
   \lambda _1^{10}+40 \lambda _1^6-24 \lambda _1^4+1) \lambda _2^6+(-5 \lambda _1^{12}+57
   \lambda _1^8-72 \lambda _1^6+24 \lambda _1^2-5) \lambda _2^4+ \\& 3 \lambda _1^2 (\lambda _1^8-4
   \lambda _1^6+8 \lambda _1^2-4) \lambda _2^2+3 \lambda _1^6-5 \lambda _1^4+1) q) \\
    \end{array}
\end{equation} }
\paragraph{Knot $5_1$}
{\small
\begin{equation}
    \begin{array}{ll}
        P_{2,3}^{5_1}(\l_1,\l_2) = & \l_1^{-8}\l_2^{-8} (\lambda _1^8 \left(\lambda _1^8-\lambda _1^6+\lambda _1^4-\lambda _1^2+1\right) \lambda _2^{16}-\lambda
   _1^6 \left(\lambda _1^{10}-2 \lambda _1^8+2 \lambda _1^6-2 \lambda _1^4+2 \lambda _1^2-1\right)
   \lambda _2^{14}+\\& \lambda _1^4 \left(\lambda _1^{12}-2 \lambda _1^{10}+\lambda _1^8-\lambda _1^6+\lambda
   _1^4-2 \lambda _1^2+1\right) \lambda _2^{12}-\lambda _1^2 \left(\lambda _1^2-1\right){}^3
   \left(\lambda _1^8+\lambda _1^6+\lambda _1^4+\lambda _1^2+1\right) \lambda _2^{10}\\& +\left(\lambda
   _1^{16}-2 \lambda _1^{14}+\lambda _1^{12}+\lambda _1^8+\lambda _1^4-2 \lambda _1^2+1\right) \lambda
   _2^8+\left(\lambda _1^2-1\right){}^3 \left(\lambda _1^8+\lambda _1^6+\lambda _1^4+\lambda
   _1^2+1\right) \lambda _2^6+\\& \left(\lambda _1^{12}-2 \lambda _1^{10}+\lambda _1^8-\lambda _1^6+\lambda
   _1^4-2 \lambda _1^2+1\right) \lambda _2^4+\left(\lambda _1^{10}-2 \lambda _1^8+2 \lambda _1^6-2
   \lambda _1^4+2 \lambda _1^2-1\right) \lambda _2^2+\\& \lambda _1^8-\lambda _1^6+\lambda _1^4-\lambda
   _1^2+1)  \\
    \end{array}
\end{equation}

\begin{equation}
    \begin{array}{ll}
        P_{3,3}^{5_1}(\l_1,\l_2) & =  -q^{-4}\l_1^{-16}\l_2^{-16} ( \lambda _1^{16} (q \lambda _1^{16}-q^2  \lambda _1^{14}+(q-2) \lambda _1^{12}+\lambda _1^{10}-q
   \lambda _1^8+q^2  \lambda _1^6+(q+1) \lambda _1^4-q^2  \lambda _1^2-1) \lambda _2^{32}-\\& \lambda
   _1^{14} (-\lambda _1^{18}+2 \lambda _1^{16}-\lambda _1^{14}-\lambda _1^{12}+2 \lambda _1^{10}-2
   \lambda _1^8-\lambda _1^6+3 \lambda _1^4+\\& q (\lambda _1^{18}-2 \lambda _1^{14}+3 \lambda _1^{12}-2
   \lambda _1^{10}+3 \lambda _1^6-\lambda _1^4-2 \lambda _1^2+1)-1) \lambda _2^{30}+\\& \lambda
   _1^{12} (-2 \lambda _1^{20}+\lambda _1^{18}+3 \lambda _1^{16}-4 \lambda _1^{14}+3 \lambda
   _1^{12}+\lambda _1^{10}-5 \lambda _1^8+2 \lambda _1^6+3 \lambda _1^4-3 \lambda _1^2+\\& q (\lambda
   _1^{20}+2 \lambda _1^{18}-5 \lambda _1^{16}+2 \lambda _1^{14}+2 \lambda _1^{12}-3 \lambda _1^{10}+3
   \lambda _1^8+2 \lambda _1^6-5 \lambda _1^4+\lambda _1^2+1)+1) \lambda _2^{28}- \\& q^2
   \lambda _1^{10} (q \lambda _1^{22}+(3-2 q) \lambda _1^{20}-2 (q+1) \lambda _1^{18}+(4 q-2)
   \lambda _1^{16}-2 (q-2) \lambda _1^{14}-(2 q+1) \lambda _1^{12}+(3 q-2) \lambda _1^{10}-\\& 2 (q-2)
   \lambda _1^8-2 (q+1) \lambda _1^6+(4 q-2) \lambda _1^4+(3-2 q) \lambda _1^2-1) \lambda
   _2^{26}-\\& \lambda _1^8 ((2 \lambda _1^{20}-3 \lambda _1^{18}-2 \lambda _1^{16}+6 \lambda
   _1^{14}-\lambda _1^{12}-3 \lambda _1^{10}+3 \lambda _1^8-3 \lambda _1^6-2 \lambda _1^4+5 \lambda
   _1^2-2) \lambda _1^2+\\& q (\lambda _1^{24}-2 \lambda _1^{22}-2 \lambda _1^{20}+4 \lambda
   _1^{18}-3 \lambda _1^{16}-2 \lambda _1^{14}+4 \lambda _1^{12}-\lambda _1^{10}-3 \lambda _1^8+4 \lambda
   _1^6-3 \lambda _1^4+1)) \lambda _2^{24}+\\& \lambda _1^6 (q^2  \lambda _1^{26}+2 \lambda
   _1^{24}+(1-3 q) \lambda _1^{22}+(q-3) \lambda _1^{20}+(2 q+1) \lambda _1^{18}+(2-4 q) \lambda
   _1^{16}+2 (q-2) \lambda _1^{14}+2 (q+1) \lambda _1^{12}+\\& (2-4 q) \lambda _1^{10}+(q-3) \lambda _1^8+(2
   q+1) \lambda _1^6+(1-3 q) \lambda _1^4+2 q^2  \lambda _1^2+1) \lambda _2^{22}+\\& \lambda _1^4
   (\lambda _1^{28}+\lambda _1^{26}-5 \lambda _1^{24}+\lambda _1^{22}+3 \lambda _1^{20}-4 \lambda
   _1^{18}+5 \lambda _1^{16}+3 \lambda _1^{14}-9 \lambda _1^{12}+2 \lambda _1^{10}+3 \lambda _1^8-3
   \lambda _1^6+3 \lambda _1^4+\lambda _1^2+\\& q (\lambda _1^{28}-3 \lambda _1^{26}+3 \lambda _1^{24}+2
   \lambda _1^{22}-4 \lambda _1^{20}+2 \lambda _1^{18}+4 \lambda _1^{16}-7 \lambda _1^{14}+5 \lambda
   _1^{12}+2 \lambda _1^{10}-4 \lambda _1^8+\lambda _1^6+2 \lambda _1^4-3 \lambda _1^2+1)-2)
   \lambda _2^{20}-\\& \lambda _1^2 (-\lambda _1^{30}+3 \lambda _1^{28}-2 \lambda _1^{26}-2 \lambda
   _1^{24}+3 \lambda _1^{22}-2 \lambda _1^{20}-3 \lambda _1^{18}+8 \lambda _1^{16}-4 \lambda _1^{14}-3
   \lambda _1^{12}+4 \lambda _1^{10}-\lambda _1^8-2 \lambda _1^6+4 \lambda _1^4-\lambda _1^2+\\& q
   (\lambda _1^{30}-\lambda _1^{28}-2 \lambda _1^{26}+4 \lambda _1^{24}-\lambda _1^{22}-2 \lambda
   _1^{20}+7 \lambda _1^{18}-4 \lambda _1^{16}-4 \lambda _1^{14}+7 \lambda _1^{12}-2 \lambda _1^{10}-2
   \lambda _1^8+4 \lambda _1^6-2 \lambda _1^4-2 \lambda _1^2+1)- \\&1)  \lambda _2^{18}+q
   (q^2  \lambda _1^{32}+2 \lambda _1^{30}-(3 q+2) \lambda _1^{28}+(4 q-2) \lambda _1^{26}-3 (q-2)
   \lambda _1^{24}-2 (q+1) \lambda _1^{22}+(9 q-4) \lambda _1^{20}-4 (q-2) \lambda _1^{18}-\\& (6 q+5)
   \lambda _1^{16}+ (8 q-4) \lambda _1^{14}+(9-5 q) \lambda _1^{12}-2 (q+1) \lambda _1^{10}+(6 q-3)
   \lambda _1^8-2 (q-2) \lambda _1^6-(3 q+2) \lambda _1^4+2 q^2  \lambda _1^2+1) \lambda
   _2^{16}-\\& q^2  (\lambda _1^{30}-(2 q+1) \lambda _1^{28}+(4 q-2) \lambda _1^{26}-2 (q-2) \lambda
   _1^{24}-(2 q+1) \lambda _1^{22}+(4 q-2) \lambda _1^{20}+(7-4 q) \lambda _1^{18}-4 (q+1) \lambda
   _1^{16}+\\& (8 q-4) \lambda _1^{14}+(7-4 q) \lambda _1^{12}-2 (q+1) \lambda _1^{10}+(3 q-2) \lambda _1^8-2
   (q-2) \lambda _1^6-2 (q+1) \lambda _1^4+(3 q-2) \lambda _1^2+1) \lambda _2^{14}+ \\& (\lambda
   _1^{28}+\lambda _1^{26}-5 \lambda _1^{24}+\lambda _1^{22}+3 \lambda _1^{20}-4 \lambda _1^{18}+5
   \lambda _1^{16}+3 \lambda _1^{14}-9 \lambda _1^{12}+2 \lambda _1^{10}+3 \lambda _1^8-3 \lambda _1^6+3
   \lambda _1^4+\lambda _1^2+\\& q (\lambda _1^{28}-3 \lambda _1^{26}+3 \lambda _1^{24}+2 \lambda
   _1^{22}-4 \lambda _1^{20}+2 \lambda _1^{18}+4 \lambda _1^{16}-7 \lambda _1^{14}+5 \lambda _1^{12}+2
   \lambda _1^{10}-4 \lambda _1^8+\lambda _1^6+2 \lambda _1^4-3 \lambda _1^2+1)-2) \lambda
   _2^{12}+\\& (q^2  \lambda _1^{26}+2 \lambda _1^{24}+(1-3 q) \lambda _1^{22}+(q-3) \lambda _1^{20}+(2
   q+1) \lambda _1^{18}+(2-4 q) \lambda _1^{16}+2 (q-2) \lambda _1^{14}+2 (q+1) \lambda _1^{12}+\\& (2-4 q)
   \lambda _1^{10}+(q-3) \lambda _1^8+(2 q+1) \lambda _1^6+(1-3 q) \lambda _1^4+2 q^2  \lambda
   _1^2+1) \lambda _2^{10}+\\& q^8 (-\lambda _1^{24}+2 \lambda _1^{22}+2 \lambda _1^{20}-4 \lambda
   _1^{18}+3 \lambda _1^{16}+2 \lambda _1^{14}-4 \lambda _1^{12}+\lambda _1^{10}+3 \lambda _1^8-4 \lambda
   _1^6+3 \lambda _1^4+\\& q (\lambda _1^{24}-5 \lambda _1^{20}+2 \lambda _1^{18}+3 \lambda _1^{16}-3
   \lambda _1^{14}+\lambda _1^{12}+2 \lambda _1^{10}-6 \lambda _1^8+2 \lambda _1^6+2 \lambda _1^4-2
   \lambda _1^2+1)-1) \lambda _2^8+\\&(\lambda _1^{22}+\lambda _1^{20}-4 \lambda _1^{18}+2
   \lambda _1^{16}+2 \lambda _1^{14}-3 \lambda _1^{12}+\lambda _1^{10}+2 \lambda _1^8-4 \lambda _1^6+2
   \lambda _1^4+\lambda _1^2+\\& q (-3 \lambda _1^{20}+2 \lambda _1^{18}+2 \lambda _1^{16}-4 \lambda
   _1^{14}+\lambda _1^{12}+2 \lambda _1^{10}-4 \lambda _1^8+2 \lambda _1^6+2 \lambda _1^4-3 \lambda
   _1^2+1)-1) \lambda _2^6+\\& (-2 \lambda _1^{20}+\lambda _1^{18}+3 \lambda _1^{16}-4
   \lambda _1^{14}+3 \lambda _1^{12}+\lambda _1^{10}-5 \lambda _1^8+2 \lambda _1^6+3 \lambda _1^4-3
   \lambda _1^2+\\& q (\lambda _1^{20}+2 \lambda _1^{18}-5 \lambda _1^{16}+2 \lambda _1^{14}+2 \lambda
   _1^{12}-3 \lambda _1^{10}+3 \lambda _1^8+2 \lambda _1^6-5 \lambda _1^4+\lambda _1^2+1)+1)
   \lambda _2^4-\\& q^2 (\lambda _1^{18}-2 q \lambda _1^{16}+(3 q-2) \lambda _1^{14}+(3-2 q) \lambda
   _1^{12}-2 \lambda _1^{10}+2 q \lambda _1^8+(3-2 q) \lambda _1^6-\\& (2 q+1) \lambda _1^4+2 q^2  \lambda
   _1^2+1) \lambda _2^2+q \lambda _1^{16}-q^2  \lambda _1^{14}+(q-2) \lambda _1^{12}+\lambda
   _1^{10}-q \lambda _1^8+q^2  \lambda _1^6+(q+1) \lambda _1^4-q^2  \lambda _1^2-1).\\
    \end{array}
\end{equation}}

\subsection{\uqslfour{}}
\paragraph{Trefoil}
{\small
\begin{equation}
    \begin{array}{ll}
         P_{2,4}^{3_1}(\l_1,\l_2,\l_3) = & \l_1^3 \l_2^4 \l_3^3 (\lambda _1^8 (\lambda _1^4-\lambda _1^2+1) \lambda _3^8 (\lambda _3^4-\lambda
   _3^2+1) \lambda _2^{16}-\lambda _1^6 (\lambda _1^6-2 \lambda _1^4+2 \lambda _1^2-1)
   \lambda _3^6 (\lambda _3^6-2 \lambda _3^4+2 \lambda _3^2-1) \lambda _2^{14}+\\& \lambda _1^4
   \lambda _3^4 (\lambda _1^8-2 \lambda _1^6+\lambda _1^4-2 \lambda _1^2+(\lambda _1^8-2 \lambda
   _1^6+\lambda _1^4-2 \lambda _1^2+1) \lambda _3^8+(-2 \lambda _1^8+5 \lambda _1^6-5 \lambda
   _1^4+5 \lambda _1^2-2) \lambda _3^6+\\& (\lambda _1^8-5 \lambda _1^6+6 \lambda _1^4-5 \lambda
   _1^2+1) \lambda _3^4+(-2 \lambda _1^8+5 \lambda _1^6-5 \lambda _1^4+5 \lambda _1^2-2)
   \lambda _3^2+1) \lambda _2^{12}+\\& \lambda _1^2 (\lambda _1^2-1) \lambda _3^2
   (\lambda _3^2-1) ((\lambda _1^6-\lambda _1^4+\lambda _1^2) \lambda
   _3^8+(\lambda _1^8-2 \lambda _1^6+\lambda _1^4-2 \lambda _1^2+1) \lambda _3^6+(-\lambda
   _1^8+\lambda _1^6+\lambda _1^2-1) \lambda _3^4+\\& (\lambda _1^8-2 \lambda _1^6+\lambda _1^4-2
   \lambda _1^2+1) \lambda _3^2+\lambda _1^6-\lambda _1^4+\lambda _1^2) \lambda
   _2^{10}+((\lambda _1^8-\lambda _1^6+\lambda _1^4) \lambda _3^{12}+\\& (\lambda
   _1^{10}-4 \lambda _1^8+5 \lambda _1^6-4 \lambda _1^4+\lambda _1^2) \lambda _3^{10}+(\lambda
   _1^{12}-4 \lambda _1^{10}+6 \lambda _1^8-4 \lambda _1^6+6 \lambda _1^4-4 \lambda _1^2+1) \lambda
   _3^8-\\& (\lambda _1^{12}-5 \lambda _1^{10}+4 \lambda _1^8+\lambda _1^6+4 \lambda _1^4-5 \lambda
   _1^2+1) \lambda _3^6+(\lambda _1^{12}-4 \lambda _1^{10}+6 \lambda _1^8-4 \lambda _1^6+6
   \lambda _1^4-4 \lambda _1^2+1) \lambda _3^4+\\& (\lambda _1^{10}-4 \lambda _1^8+5 \lambda _1^6-4
   \lambda _1^4+\lambda _1^2) \lambda _3^2+\lambda _1^8-\lambda _1^6+\lambda _1^4) \lambda
   _2^8+(\lambda _1^2-1) (\lambda _3^2-1) ((\lambda _1^6-\lambda
   _1^4+\lambda _1^2) \lambda _3^8+\\& (\lambda _1^8-2 \lambda _1^6+\lambda _1^4-2 \lambda
   _1^2+1) \lambda _3^6+(-\lambda _1^8+\lambda _1^6+\lambda _1^2-1) \lambda
   _3^4+(\lambda _1^8-2 \lambda _1^6+\lambda _1^4-2 \lambda _1^2+1) \lambda _3^2+\\& \lambda
   _1^6-\lambda _1^4+\lambda _1^2) \lambda _2^6+(\lambda _1^8-2 \lambda _1^6+\lambda _1^4-2
   \lambda _1^2+(\lambda _1^8-2 \lambda _1^6+\lambda _1^4-2 \lambda _1^2+1) \lambda
   _3^8+\\& (-2 \lambda _1^8+5 \lambda _1^6-5 \lambda _1^4+5 \lambda _1^2-2) \lambda
   _3^6+(\lambda _1^8-5 \lambda _1^6+6 \lambda _1^4-5 \lambda _1^2+1) \lambda _3^4+\\& (-2
   \lambda _1^8+5 \lambda _1^6-5 \lambda _1^4+5 \lambda _1^2-2) \lambda _3^2+1) \lambda
   _2^4-(\lambda _1^6-2 \lambda _1^4+2 \lambda _1^2-1) (\lambda _3^6-2 \lambda _3^4+2
   \lambda _3^2-1) \lambda _2^2+\\& (\lambda _1^4-\lambda _1^2+1) (\lambda _3^4-\lambda
   _3^2+1)). \\
    \end{array}
\end{equation}}

\section{Appendix B. Invariants $\P_{m,N}^{\L}$ of links \label{appB}}
Here we listed the unreduced invariants of links, corresponding to the nilpotent representations with parameters $W_{m,N}$ for different $m$ and $N$. We worked out the following links: the Hopf link --- link $L_2a_1$, the Whitehead link --- link $L_5a_1$, the Borromean rings --- link $L_6 a_4$, and link $L_7 a_1$.

The invariants depend on a set of parameters $\l_1^{(i)}, \dots, \l_{N-1}^{(i)}$, where $(i)$ enumerates different colors on different strands of links. In order to simplify the expressions of the invariants below, we use here the following notation: $\l_1^{(i)} = \a_i$, $\l_2^{(i)} = \b_i$, $\l_3^{(i)} = \g_i$.

\subsection{Hopf link}
The invariants of the Hopf link ($L_2 a_1$) do not depend on parameters $\l_i$ of representations $W_{m,N}$ and are combinations of the framing factor $q^{C_2^{\mathcal{U}_q(sl_N)}}$ in representation $R_{m,N} = [(N-1)(m-1), \dots, m-1]$ and the normalization coefficient $\Xi_{m,N}$ (\ref{norm}) at a special point $\l_i=q^{m-1}$.

\begin{align}
    \P_{m,N}^{L_2a_1} & =  (-1)^{(m-1)\left(\frac{N^2(N-1)}{2}+\delta_{2,N}\right)}\,\, q^{-\frac{1}{6}N(N^2-1)(m^2-1)}q^{-\frac{1}{4}N(N-1)\,m(m-1) } \\ & =  q^{-2 C_2^{\mathcal{U}_q(sl_N)}}\frac{m^{N(N-1)/2}}{\Xi_{m,N}(\lambda=q^{m-1})},
\end{align}
where $\delta_{2,N}$ is the Kronecker symbol, $C_2^{\mathcal{U}_q(sl_N)}$ is the quadratic Casimir operator of $\mathcal{U}_q(sl_N)$, and
\begin{align}
    C_2^{\mathcal{U}_q(sl_N)}(R_{m,N}) & = \frac{1}{12} N(N^2-1)(m^2-1) \\
    {\rm dim}(R_{m,N}) & = m^{N(N-1)/2} \\
    \Xi_{m,N}(\lambda=q^{m-1}) & =  (-1)^{(m-1)\frac{N^2(N-1)}{2}}\,m^{N(N-1)/2}\,q^{\frac{1}{4}N(N-1)\,m(m-1) }
\end{align}

\subsection{More links}
\subsubsection{\uqsltwo{}}
\begin{equation}
    \begin{array}{ll}
          \P_{2,2}^{L_5 a_1} = &   \a^{-1} \b^{-1} ( \a^{2}-1)( \b^{2}-1) ,\\
          \P_{2,2}^{L_6 a_4} = &  \a^{-1} \b^{-1}\g^{-1}( \a^{2}-1)( \b^{2}-1)(\g^{2}-1)   ,\\
          \P_{2,2}^{L_7 a_1} = &  \a^{-3} \b^{-1} ( \a^{2}-1) ( \a^{4}-\a^{2}+1)( \b^{2}-1)  .\\
    \end{array}
\end{equation}

\subsubsection{\uqslthree{}}
\begin{equation}
    \begin{array}{ll}
        \P_{2,3}^{L_5 a_1} = & -q\, \a_1^{-2}\a_2^{-2}\b_1^{-2}\b_2^{-2} (\alpha _1^2 (\alpha _1^2-1) (\alpha _2^2-1) (\beta _2^2-1) \beta _1^4 (\alpha _2^2 \beta _2^2+1)+\\& \beta _1^2 (-(\alpha _1^2-1){}^2 \alpha _2^2 (\alpha _2^2-1) \beta
   _2^4+((\alpha _2^2-1){}^2 \alpha _1^4-2 (\alpha _2^4+1) \alpha _1^2+(\alpha _2^2-1){}^2) \beta _2^2+ (\alpha _1^2-1){}^2 (\alpha _2^2-1))+\\&(1-\alpha _1^2) (\alpha
   _2^2-1) (\beta _2^2-1) (\alpha _2^2 \beta _2^2+1)) , \\
        \P_{2,3}^{L_6 a_4} = & -q\, \a_1^{-2}\a_2^{-2}\a_3^{-2}\b_1^{-2}\b_2^{-2}\b_3^{-2} (\alpha _1^2 (\alpha _1^2-1) (\alpha _2^2-1) (\alpha _3^2-1) (\beta _2^2-1) (\beta _3^2-1) \beta _1^4 (\alpha _2^2 \beta _2^2+1) (\alpha _3^2 \beta _3^2+1)-\\& \beta _1^2
   ((\alpha _1^2-1){}^2 \alpha _2^2 (\alpha _2^2-1) (\alpha _3^2-1) (\beta _3^2-1) \beta _2^4 (\alpha _3^2 \beta _3^2+1)+ \beta _2^2 (-(\alpha _1^2-1){}^2 (\alpha
   _2^2-1){}^2 \alpha _3^2 (\alpha _3^2-1) \beta _3^4+\\&((\alpha _2^2-1){}^2 (\alpha _3^2-1){}^2 \alpha _1^4-2 ((\alpha _3^2-1){}^2 \alpha _2^4-2 (\alpha _3^4-6 \alpha _3^2+1) \alpha
   _2^2+(\alpha _3^2-1){}^2) \alpha _1^2+(\alpha _2^2-1){}^2 (\alpha _3^2-1){}^2) \beta _3^2+\\& (\alpha _1^2-1){}^2 (\alpha _2^2-1){}^2 (\alpha _3^2-1))-(\alpha
   _1^2-1){}^2 (\alpha _2^2-1) (\alpha _3^2-1) (\beta _3^2-1) (\alpha _3^2 \beta _3^2+1))+\\&(1-\alpha _1^2) (\alpha _2^2-1) (\alpha _3^2-1) (\beta _2^2-1)
   (\beta _3^2-1) (\alpha _2^2 \beta _2^2+1) (\alpha _3^2 \beta _3^2+1)) ,
   \end{array}
   \end{equation}
   \begin{equation}
   \begin{array}{ll}
        \P_{2,3}^{L_7 a_1} = &  -q\, \a_1^{-6}\a_2^{-2}\b_1^{-6}\b_2^{-2} (\alpha _1^6 (\alpha _1^6-2 \alpha _1^4+2 \alpha _1^2-1) (\alpha _2^2-1) (\beta _2^2-1) (\alpha _2^2 \beta _2^2+1) \beta _1^{12}-\\& 2 \alpha _1^4 (\alpha _1^2-1){}^2 (\alpha _1^4-\alpha
   _1^2+1) (\alpha _2^2-1) (\beta _2^2-1) (\alpha _2^2 \beta _2^2+1) \beta _1^{10}+\\& 2 \alpha _1^2 (\alpha _1^2-1) ((\alpha _1^8-3 \alpha _1^6+3 \alpha _1^4-3 \alpha _1^2+1) \alpha _2^2
   (\alpha _2^2-1) \beta _2^4-\\& ((\alpha _2^2-1){}^2 \alpha _1^8-3 (\alpha _2^2-1){}^2 \alpha _1^6+(3 \alpha _2^4-8 \alpha _2^2+3) \alpha _1^4-3 (\alpha _2^2-1){}^2 \alpha _1^2+(\alpha
   _2^2-1){}^2) \beta _2^2+\\& (-\alpha _1^8+3 \alpha _1^6-3 \alpha _1^4+3 \alpha _1^2-1) (\alpha _2^2-1)) \beta _1^8+(-(\alpha _1^2-1){}^2 (\alpha _1^8-4 \alpha _1^6+3 \alpha _1^4-4 \alpha
   _1^2+1) \alpha _2^2 (\alpha _2^2-1) \beta _2^4+\\& ((\alpha _2^2-1){}^2 \alpha _1^{12}-6 (\alpha _2^2-1){}^2 \alpha _1^{10}+4 (3 \alpha _2^4-7 \alpha _2^2+3) \alpha _1^8-2 (7 \alpha _2^4-12
   \alpha _2^2+7) \alpha _1^6+4 (3 \alpha _2^4-7 \alpha _2^2+3) \alpha _1^4-\\& 6 (\alpha _2^2-1){}^2 \alpha _1^2+(\alpha _2^2-1){}^2) \beta _2^2+(\alpha _1^2-1){}^2 (\alpha _1^8-4 \alpha _1^6+3
   \alpha _1^4-4 \alpha _1^2+1) (\alpha _2^2-1)) \beta _1^6-\\& 2 (\alpha _1^2-1) ((\alpha _1^8-3 \alpha _1^6+3 \alpha _1^4-3 \alpha _1^2+1) \alpha _2^2 (\alpha _2^2-1) \beta
   _2^4-((\alpha _2^2-1){}^2 \alpha _1^8-3 (\alpha _2^2-1){}^2 \alpha _1^6+\\& (3 \alpha _2^4-8 \alpha _2^2+3) \alpha _1^4-3 (\alpha _2^2-1){}^2 \alpha _1^2+(\alpha _2^2-1){}^2) \beta
   _2^2+(-\alpha _1^8+3 \alpha _1^6-3 \alpha _1^4+3 \alpha _1^2-1) (\alpha _2^2-1)) \beta _1^4-\\& 2 (\alpha _1^2-1){}^2 (\alpha _1^4-\alpha _1^2+1) (\alpha _2^2-1) (\beta _2^2-1)
   (\alpha _2^2 \beta _2^2+1) \beta _1^2+\\& (-\alpha _1^6+2 \alpha _1^4-2 \alpha _1^2+1) (\alpha _2^2-1) (\beta _2^2-1) (\alpha _2^2 \beta _2^2+1)) , \\
    \end{array}
\end{equation}

\section{Appendix C. Alexander polynomials of knots and links \label{appC}}
We list here the Alexander polynomials of knots and links that we mention in the paper. The variables are those used throughout the paper. These answers coincide with expressions from \cite{katlas} after the change of variables ($t \rightarrow \l^2$).

\begin{align}
    \A^{3_1}(\l) & = \l^2 - 1 + \l^{-2},  \nn \\
    \A^{4_1}(\l) & = -\l^2 + 3 - \l^{-2}, \nn  \\
    \A^{4_1}(\l) & =\l^4-  \l^2 + 1- \l^{-2}+\l^{-4},  \nn
\end{align}
\begin{align}
  &  \D^{L_2 a_1}  = -1, \nn \\
  &  \D^{L_5 a_1}(\l,\m) = \frac{(\l^2-1)(\m^2 -1)}{\l \m}, \nn \\
  & \D^{L_6 a_4}(\l,\m,\n) = \frac{(\l^2-1)(\m^2-1)(\n^2-1)}{\l \m \n}, \nn \\
  & \D^{L_7 a_1} (\l,\m)=\frac{(\l^2-1)(\m^2-1)(\l^2-\l+1)}{\l^3 \m}.\nn
  \end{align}

\end{document}